\documentclass[11pt,twoside]{article}   
\usepackage{amsmath,epsfig,graphicx,amssymb}

\usepackage{epsf}

\def\simt0{\mathrel{\mathop{\sim}\limits_{{}^{\tau\to 0}}}}
\def\be{\begin{equation}}
\def\ee{\end{equation}}
\def\bea{\begin{eqnarray}}
\def\eea{\end{eqnarray}}

\date{}
\title{Bosonization of the Pairing Hamiltonian}
\author{ M.B. Barbaro, M.R. Quaglia
\\
Dipartimento di Fisica Teorica, Universit\`a 
di Torino\\
and\\
INFN, Sezione di Torino,
Via P. Giuria 1, I-10125, Turin, Italy}
\begin{document}
\maketitle


\begin{minipage}{5in}
\centerline{{\sc Abstract}}
\medskip
We address the problem of the bosonization of finite fermionic systems 
with two different approaches. 

First we work in the path integral formalism, showing how a truly bosonic 
effective action can be derived from a generic fermionic one with a quartic 
interaction. We then apply our scheme to the pairing hamiltonian in the 
degenerate case proving that, in this instance, 
several of the features characterizing the spontaneous breaking of the 
global gauge symmetry U(1) occurring in the infinite system persist in the
finite system as well.  

Accordingly we interpret the excitations associated with the addition and 
removal of pairs of fermions as a quasi-Goldstone boson and the excitations
corresponding to the breaking of a pair (seniority one states in the language 
of the pairing hamiltonian) as Higgs modes.  

Second, we face the more involved problem of a non-degenerate single particle
spectrum, where one more kind of excitations arises, corresponding to the 
promotion of pairs to higher levels. This we do by solving directly the 
Richardson equations. From this analysis 
the existence emerges 
of critical values of the coupling constant, which signal the 
transition between two regimes, one dominated by the mean field physics, 
the other by the pairing interaction.
\end{minipage}

\section{Introduction}

The problem of the bosonization of finite and infinite fermionic systems has 
been extensively explored in the past. All these investigations have been prompted
by the recognition that in nature this phenomenon is widely occurring. 
Best examples of it are offered by the superconductivity in metals with the 
related BCS theory \cite{Bardeen:1957kj} and by the superfluidity of 
certain atomic nuclei 
\cite{BMP} where it is signalled by the existence of a gap (about 1 MeV) in 
the energy spectrum, by the reduction of about a factor of two in the moment 
of inertia with respect to the shell model prediction and by the staggering 
behavior of the separation energies (odd-even effect in the mass number A). 
More generally in the nuclear case the bosonization phenomenon is of course 
epitomized by the outstandingly successful Interacting Boson Model (IBM) 
by Arima and Iachello \cite{Iac}.
In this paper we first address the problem of the bosonization in the path 
integral formalism. Although this approach is general and hence keeps its 
validity in bosonizing {\it any} fermionic action with a quartic interaction, 
for sake of illustration we shall consider the well-known pairing hamiltonian 
which reads   
\begin{equation}
\label{eq:HP}
\hat H\equiv \hat H_0+\hat H_P
=\sum_{\nu=1}^L e_\nu \sum_{m_\nu=-j_\nu}^{j_\nu}
\hat a^\dagger_{j_\nu m_\nu} \hat a_{j_\nu m_\nu}
- G \sum_{\mu,\nu=1}^L \hat A^\dagger_\mu \hat A_\nu~.
\end{equation}
In \eqref{eq:HP} $\hat a_{j_\nu m_\nu}$ is the destruction 
operator of a fermion in a single particle level (whose number is $L$)
characterized by angular momentum $j_\nu$, third component $m_\nu$
and single particle energy $e_\nu$, whereas the operator 
\begin{equation}
\hat A_\nu= \sum_{m_\nu=1/2}^{j_\nu} (-1)^{j_\nu-m_\nu}
\hat a_{j_\nu,- m_\nu} \hat a_{j_\nu m_\nu}
\label{eq:A}
\end{equation} 
destroys a pair of fermions with total angular momentum $J=0$ in the level 
$j_\nu$.

The Hamiltonian \eqref{eq:HP} is indeed:
\begin{enumerate}
\item
relevant for both nuclear and condensed matter physics. 
In nuclear physics it is known to represent, together with the quadrupole 
force,
a dominant part of the residual nucleon-nucleon interaction and 
a crucial microscopic ingredient of the IBM~\cite{Iac}.
If the interacting fermions are electrons, $\hat H$ (in particular $\hat H_P$) 
is the Hamiltonian
underlying the BCS theory of 
superconductivity~\cite{Bardeen:1957kj,Bardeen:1957mv}. Moreover, 
the same Hamiltonian
governs collective phenomena in the physics of liquids and metal clusters;
\item
a simple example to illustrate the bosonization of fermionic 
systems.
The problem of relating the {\em fermionic} Hamiltonian (\ref{eq:HP})
to a {\em bosonic} one, like the IBM, with the same spectrum,
has been the object of several investigations\cite{Gin,Cas,Iac}
and still represents a challenging question. 

Furthermore:
\item
for infinite systems, the BCS theory, of which the hamiltonian \eqref{eq:HP}
offers a simple realization, is an example of
spontaneous symmetry breaking of the gauge group $U(1)$, with the 
associated appearance of a Goldstone boson~\cite{Wei}, set up
with excitations consisting of the addition and removal of pairs of particles
to and from the system.
The question then arises whether this Goldstone mode survives in finite 
systems and, if so, how the Goldstone and Higgs excitations should be 
identified.
\item
The Hamiltonian \eqref{eq:HP} has been shown to be exactly 
integrable~\cite{Cambiaggio:1997vz}, which means that the number of
constants of motion equals the number of degrees of freedom.
This does not imply, however, that explicit expressions for its 
eigenvalues and eigenstates are easily obtained, since the constants 
of motion are usually complicated
operators which, in general, can only be diagonalized numerically. 
For this reason it is useful to provide approximate analytic
expressions for the eigenvalues and eigenvectors of $\hat H_P$.
\end{enumerate}

This report is organized as follows.
In Section \ref{sec:deg} we illustrate how the path integrals scheme works and 
apply it to \eqref{eq:HP} in the degenerate case, where only one single 
particle level is considered, deducing the well-known zero seniority spectrum 
in the path integral formalism, discussing the appearance of a Goldstone boson
and studying the seniority excitations within a matrix approach in 
Sec.~\ref{sec:matrix}. 
In Section~\ref{sec:nondeg} we study, by directly solving the Richardson 
equations, the problem of one and two pairs living in many
non-degenerate single particle levels, deriving analytic expressions 
for the energies of both the collective and the trapped states. 
Finally, in Sec.~\ref{sec:BCS} we briefly discuss the 
relation between the exact and BCS solution to the pairing problem  in terms 
of Bogolioubov quasi-particles.

\section{The degenerate case}
\label{sec:deg}

The pairing Hamiltonian has been first diagonalized by Kerman, Lawson and 
MacFarlane~\cite{Ker61} in the simple case of one degenerate single
particle level 
by using group theoretical methods. In fact $\hat H$ can be cast in the
following form
\begin{equation}
\hat H=2\sum_{\nu=1}^L e_\nu\hat S_{z\nu}+
\sum_{\nu=1}^L e_\nu\Omega_\nu-G\hat S_+\hat S_-\ ,
\end{equation}
$\Omega_\nu=j_\nu+1/2$ being the pair degeneracy of the level $j_\nu$,
in terms of the quasi-spin operators~\cite{Ker61a,And58}
\begin{eqnarray}
&&\hat S_+=\sum_{\nu=1}^L \hat A^\dagger_\nu\ ,
\ \ \ \ \ \ \ 
\hat S_-=\sum_{\nu=1}^L \hat A_\nu\ ,
\\
&&\hat S_z=\frac{1}{2}\sum_{\nu=1}^L \hat S_{z\nu}
=\sum_{\nu=1}^L\sum_{m_\nu=1/2}^{j_\nu} 
\left(\hat a^\dagger_{j_\nu m_\nu}\hat a_{j_\nu m_\nu}
-\hat a_{j_\nu -m_\nu}\hat a^\dagger_{j_\nu -m_\nu}
\right)
\end{eqnarray}
which span a SU(2) algebra.
Hence the pure pairing Hamiltonian $\hat H_P$ can be immediately diagonalized,
yielding
\begin{equation}
\label{eq:6}
-G <\hat S_+ \hat S_->=-G \left[N/2\left(\Omega-N/2+1\right)
-v/2\left(\Omega-v/2+1\right)\right]~,
\end{equation}
where $N$ is the eigenvalue of the particle number operator 
(the number of fermions is assumed here to be an even integer)
\begin{equation}
\hat N=\sum_{\nu=1}^L\sum_{m_\nu=-j_\nu}^{j_\nu}\hat a^\dagger_{j_\nu m_\nu}\hat a_{j_\nu m_\nu}
\end{equation} 
and $\Omega=\sum_\nu\Omega_\nu$ the total degeneracy.
The seniority quantum number 
\begin{equation}
v=\Omega-2 S~,
\end{equation}
$S(S+1)$ being the eigenvalue of $\hat S^2$,
represents the number of unpaired particles.

Now, in the degenerate case $(L=1)$ $e_\nu=e$ and the full Hamiltonian becomes
\begin{equation}
\hat H=2 e \,(\hat S_z+\Omega)-G\hat S_+\hat S_-\ .
\end{equation}
Hence the spectrum generating algebra is still SU(2) and the energy turns out 
to read
\begin{equation}
\label{eq:E}
E(n,s) = 2 e n-G \left[n\left(\Omega-n+1\right)
-s\left(\Omega-s+1\right)\right]~.
\end{equation}
In \eqref{eq:E}, for future convenience, the pair number 
$n=N/2\le\Omega$
and the quantum number 
\begin{equation}
\label{eq:s}
s=v/2\le\min\{n,\Omega-n\}~,
\end{equation}
which counts the number of broken {\it pairs}, have been introduced.

The spectrum (\ref{eq:E}) is associated with two independent types of excitations:
one is related to the addition or removal of one pair of fermions and the
other to the breaking of a pair.
The former, described by the quantum number $n$, is a Goldstone boson 
associated with the spontaneous 
breaking of the global gauge invariance reflecting the particle number
conservation; the latter, described by the quantum number $s$, 
can be viewed as corresponding to the Higgs excitations~\cite{Barbaro:2004nk}.

 In an infinite system the energy of the Goldstone bosons vanishes with
the associated quantum number. This does not occur in a finite system, but, to
the extent that the energy spectrum of the latter displays a pattern 
similar to that of an infinite system, it should exhibit two quite different 
energy scales.
Actually the excitation energies associated with both the quantum 
numbers $n$ and $s$ appear to be of the same order, namely $g\Omega$.  
However the Goldstone nature of the energy spectrum associated with
$n$ is clearly apparent when one considers the excitations 
with respect to the minimum of \eqref{eq:E}. This occurs for 
\be 
n=n_0=[\nu_0], 
\ee 
$[...]$ meaning integer part, with  
\be 
\nu_0\equiv \frac{1}{2}(\Omega + 1) - \frac{e}{G}    
\,. 
\label{n0} 
\ee 
Introducing then the shifted quantum number 
\be 
\nu=n-n_0 \,,
\label{nu} 
\ee 
 (\ref{eq:E}) becomes 
\be 
E(n_0+\nu,s) = G\nu^2 + 2G \nu (n_0 - \nu_0 ) - G n_0 ( 2\nu_0 - n_0 )    
+G s(\Omega-s+1)\,. 
\label{eq:exact} 
\ee 
Therefore the addition (or removal) of one pair of nucleons with respect to 
the ground state requires an energy of order $G$: this is the energy of the 
Goldstone boson. The energy required to break a pair, namely the seniority 
energy, is instead of order $G\,\Omega$: this is the energy of a Higgs boson. 

\subsection{The path integral approach}


In this Section we show that the pairing spectrum (\ref{eq:E}) 
for $s=0$ can be obtained in the framework of the Feynman path integral and
illustrate how an effective bosonic action can be set up in this case.

To do this it is necessary to introduce the odd Grassmann variables 
$\lambda_m$, $\lambda_m^*$
associated to the fermionic destruction and creation operator $\hat a_m$,
$\hat a^\dagger_m$ (we omit the index $j$ since we are considering only
one single particle level) and the even elements
\begin{equation}
\varphi_m(t)=(-1)^{j-m}\lambda_{-m}(t)\lambda_m(t),\;\;\;
\varphi^*_m(t)=(-1)^{j-m}\lambda^*_m(t)
\lambda^*_{-m}(t)~,
\label{eq:a}
\end{equation}
describing pairs of fermions with $J_z=0$. 

Since we restrict ourselves to the zero seniority part 
of the spectrum (\ref{eq:E}), we define 
\begin{equation}
\label{eq:Phi}
\Phi(t)=\sum_{m>0} (-1)^{j-m} \lambda_{-m}(t) \lambda_m(t)~,
\end{equation}
namely the even 
elements of the Grassmann algebra associated to the $J=0$ composites 
(\ref{eq:A}), also referred to as ``hard bosons''. 
The analysis of seniority excitations, which will be studied in \ref{sec:matrix} in a different framework, would clearly require the introduction
of $J\ne 0$ composites.

Note that the index of nilpotency of the composites (\ref{eq:Phi}) is $\Omega$
(i.e., $\Phi^n=0$, $\forall n>\Omega$), reflecting the Pauli principle.

In terms of the above variables the euclidean action associated to the Hamiltonian
(\ref{eq:HP}) reads
\begin{equation}
\label{eq:S}
S=\tau\sum_{t=-N_0/2}^{N_0/2-1} 
\left\{\sum_m \lambda^*_m(t)
(\nabla^+_t+e)\lambda_m(t-1)- G \Phi^*(t) \Phi(t-1)\right\}~,
\end{equation}
where the time has been discretized ($\tau$ is the time spacing), $N_0$ is
the number of sites of the lattice,
$\nabla^+_t$ the discrete time derivative
\begin{equation}
\nabla^\pm_t f(t)=\pm\frac{1}{\tau}[f(t\pm 1)-f(t)]
\end{equation}
and $e$  the single particle fermion energy.
The variables $\lambda_m$ at the time $t=-N_0/2$ are related to the ones at
$t=N_0/2-1$ by antiperiodic boundary conditions
\begin{equation}
\lambda_m(-N_0/2)=-\lambda_m(N_0/2-1)\;.
\label{eq:antip}
\end{equation}
The action (\ref{eq:S}) can be written in terms of the even variables as
\begin{equation}
S = \sum_{t=-N_0/2}^{N_0/2-1} 
\left\{\sum_{m>0} \left[-\varphi^*_m(t) \varphi_m(t) - x^2 \varphi^*_m(t)
\varphi_m (t-1)\right]  
- G\tau \Phi^*(t) \Phi(t-1)\right\}
\label{eq:Sp},
\end{equation}
where
\begin{equation}
x=1-\tau e\;.
\end{equation}

Following the approach of Ref.~\cite{Palu} for performing Berezin 
integrals over composite variables,
the generating functional and the correlation functions of $n$ spin zero pairs 
of nucleons can then be expressed as integrals over the 
$\varphi$ variables of the composites, according to
\begin{equation}
Z=\int [d\varphi^*d\varphi]e^{-S}
\end{equation}
and 
\begin{equation}
< \Phi^n(t_2) \left[\Phi^*(t_1)\right]^n >
= \frac{1}{Z} \int [d\varphi^*d\varphi] \Phi^n(t_2) \left[\Phi^*(t_1)\right]^n
e^{-S}  \ ,
\end{equation}
respectively.
In the continuum limit the following result is then obtained~\cite{Barbaro:1996pc}
\begin{equation}
< \Phi^n(t_2) \left[\Phi^*(t_1)\right]^n > \simt0
e^{-(\beta_2-\beta_1) E_n}
\end{equation}
for the correlation function, with $\beta_{1,2}=\tau t_{1,2}$ and
\begin{equation}
E_n=n[2e-G(\Omega-n+1)]
\label{eq:spectrum}
\end{equation}
for the energy spectrum. 
The latter exactly coincides with (\ref{eq:E}) when $s=0$:
hence the well-known zero-seniority formula for the spectrum of the system 
is recovered.

Concerning the bosonization of the fermionic system,
although in the pairing Hamiltonian the fermionic and bosonic degrees of 
freedom get mixed, a purely bosonic effective 
action for the field $\Phi$ can be deduced using the variables
above introduced. It reads~\cite{Barbaro:1996pc}
\begin{equation}
S_{eff}^{(\Phi)} = \sum_{t=-N_0/2}^{N_0/2-1}  \left\{
\sum_{k=1}^\Omega \alpha_k 
\left[\Phi^*(t)\right]^k \left[ \Phi^k(t) + x^{2k} \Phi^k(t-1) \right]
- G\tau \Phi^*(t)\Phi(t-1)\right\}\ ,
\label{eq:eff}
\end{equation}
$\alpha_k$ being numerical coefficients whose asymptotic behavior is
found to be
\begin{equation}
\alpha_k
\mathrel{\mathop{\sim}\limits_{{}^{\Omega\to\infty}}}
\Omega^{-(2k-1)}~.
\end{equation} 
The action $S_{eff}^{(\Phi)}$ is equivalent
to the fermionic action (\ref{eq:S}) in the sense that it leads to the same 
spectrum (\ref{eq:E}).
Worth noticing is that 
its structure is characterized by the absence of terms of the type 
$ [\Phi^*(t)]^k [\Phi(t)]^{k-p} [\Phi(t-1)]^p $ ($p\neq 0$), due to the
occurrence of nontrivial cancellations.

\subsection{Goldstone bosons in the Hubbard-Stratonovitch approach}

In what follows we again consider the part of the spectrum 
(\ref{eq:E}) associated with the addition and removal of pairs, 
showing its connection with a Goldstone boson.

 It is well-known that the spontaneous breaking of the {\it local}
electromagnetic gauge invariance plays a crucial role in the theory of 
superconductivity~\cite{And63,Wei,Fubini:1992mr}.
Here the symmetry group U(1) of the {\it global} gauge transformations
$\psi\to e^{i\alpha}\psi$ is broken by the non-vanishing expectation value
of pair operators (the Cooper pairs), which carry charge $-2$: 
the symmetry group is thus reduced to
the unbroken subgroup $Z_2$, consisting of the two gauge transformations with 
$\alpha=0$ and $\pi$.
The  Goldstone field associated to this symmetry breaking lives in the
coset space U(1)/$Z_2$ and displays only derivative interactions.

 In a finite system, like the one considered here, a spontaneous symmetry
breaking cannot rigorously occur, since tunnelling takes place between the 
various possibly degenerate states and the true ground state turns out to 
be a unique linear superposition of the degenerate states~\cite{Wei}. 
However, if the two above mentioned features of the Goldstone fields 
in infinite systems, namely that they parameterize the coset space U(1)/$Z_2$ 
and have only derivative interactions, 
survive in finite systems, then the identification of a Goldstone boson 
becomes of 
relevance not only for a deeper understanding of the bosonization mechanism, 
but also for a convenient choice of the variables.

To explore this occurrence we perform a Hubbard-Stratonovitch linearization 
of the action (\ref{eq:S}), introducing an integration over
auxiliary fields in the generating functional
\begin{equation}
Z = \int [d\lambda d\lambda^* d\eta d\eta^*] e^{-S}~,
\end{equation}
where the new action is
\begin{eqnarray}
S &=& \tau \sum_{t=-N_0/2}^{N_0/2-1}  \Bigg\{  
G \left[\eta^*(t) \eta(t) +  \eta^* (t) \Phi(t) + \eta(t) \Phi^*(t) \right] 
\nonumber\\ 
&+& 
\sum_{m=-j}^j 
\left[ \lambda^*_m (t)  
\left( \nabla_t^+ +e\right) \lambda_m(t) \right] 
\Bigg\}
\label{SII} 
\end{eqnarray}
and the auxiliary bosonic fields $\eta^*$ and $\eta$ satisfy 
periodic boundary conditions. 
 
Next we introduce the 
following polar representation~\cite{Wei} for the auxiliary fields  
\begin{eqnarray} 
\eta &=& \sqrt{\rho} e^{2i\theta},  \,\,\,\,\, \eta^* = \sqrt{\rho} 
 e^{-2i\theta}
\label{eq:polar} 
\end{eqnarray} 
and explore whether the field $\theta$ can be identified with the 
Goldstone boson.

First, for the change of variable (\ref{eq:polar}) to be  one to one 
(with the only exception of the point $\rho=0$), $\theta$ must vary in the 
range $0 \le \theta < \pi$: hence the field $\theta$ 
lives in the coset space of the broken symmetry group $U(1)$  
of particle conservation with respect  to the unbroken subgroup $Z_2$, 
as appropriate to a Goldstone field~\cite{Wei}. 
 
Second, the Goldstone field should display
only derivative couplings in the action, which is not the case
for the field $\theta$ after the transformation (\ref{eq:polar}).
However the non-derivative coupling can be eliminated 
introducing the following transformation on the nucleon fields:  
\be 
\lambda_m = e^{i\theta} \psi_m,   
\,\,\,\,\, \lambda^*_m = e^{-i\theta}  
\psi^*_m\,. 
\label{transf} 
\ee 
In fact, as a consequence of the above, the following operators 
\begin{eqnarray} 
q^{\pm} 
 &=& \exp ( \mp i \theta) \nabla_t^{\pm} \exp ( \pm i \theta) \pm e ~,
\end{eqnarray} 
whose time matrix elements are
\begin{eqnarray} 
\left(q^+\right)_{t_1 t_2} &=& \frac{1}{\tau}\left[ 
\exp\left\{i\tau\left(\nabla_t^+\theta\right)_{t_1}\right\}\delta_{t_2,t_1+1} 
-\delta_{t_1 t_2}\right]+e \delta_{t_1 t_2} 
\\ 
\left(q^-\right)_{t_1 t_2} &=& \frac{1}{\tau}\left[\delta_{t_1 t_2}- 
\exp\left\{i\tau\left(\nabla_t^+\theta\right)_{t_2}\right\}\delta_{t_2,t_1-1} 
\right]-e \delta_{t_1 t_2}\,, 
\end{eqnarray} 
will appear in the action. In the latter 
the $\theta$ field appears only in these operators and therefore  
under derivative, as appropriate to a Goldstone field. 
 
\subsection{The saddle point expansion}

Having proved the Goldstone nature of the excitations associated to the addition or removal of fermionic pairs, 
we now proceed to deduce the energy spectrum through the saddle point expansion.

After integrating over the fermionic fields $\psi_m$ and $\psi^*_m$
we get for the generating functional the expression 
\be 
Z=\int_0^\infty \left[d\rho \right] \int_0^{\pi} \left[ d\theta \right] \exp (-S_{eff})\,, 
\label{Zeff} 
\ee 
with 
\be 
\label{eq:Seff}
S_{eff} = \tau \sum_t G \rho  - \mbox{Tr} \ln \left(-q^-q^+ +G^2\rho\right)\ ,
\ee 
where the trace is meant to be taken over the quantum number $m>0$ 
and the time. 
The $U(1)$ symmetry is now realized 
in the invariance of $S_{eff}$ under the substitution 
\be 
\theta \rightarrow \theta+\alpha\,, 
\label{invariance} 
\ee 
with $\alpha$ time independent.  

To perform the saddle point expansion one should 
first look for a minimum of the effective action (\ref{eq:Seff}) 
at constant fields.
Denoting with $\overline\rho$ the time-independent component of the $\rho$ field and defining
\begin{equation}
M = \sqrt{e^2+G ^2 \overline \rho} 
\label{eqM} 
\end{equation}
and
\begin{equation}
P^{-1} = -(1-e \tau)\  \nabla_t^+\nabla_t^- + M^2\,, 
\label{P-1} 
\end{equation}
we can write $S_{eff}$ at constant fields as follows
\begin{equation}
\overline S_{eff} =
\tau \sum_t G \overline \rho - \mbox{Tr} \ln P^{-1}~.
\end{equation}
By performing the trace over $m$ and converting the sum into a time 
integral, taking first the $N_0\to\infty$  limit and then letting
$\tau\to 0$ (with $\tau N_0$ constant), the above
becomes~\cite{Barbaro:2004nk} 
\begin{equation}
\overline S_{eff} 
= \tau N_0 \left(G\overline\rho+\Omega e-\Omega M\right) ~,
\end{equation}
whose minimum with respect to $\overline\rho$ occurs when
\begin{equation}
M=\frac{G\Omega}{2}~, 
\label{Mmin}
\end{equation}
or, equivalently, at
\begin{equation}
\overline\rho=\overline\rho_0\equiv 
\frac{1}{(2G)2}\left[(G\Omega)^2-4 e^2\right]=\frac{\Delta^2}{G^2}\,.
\label{ovrho0} 
\end{equation}
It is remarkable that the dimensionless $\overline\rho_0$, when multiplied
by $G^2$, coincides with the well-known gap $\Delta$ characterizing the BCS
theory in the one single particle level case.
The action $\overline S_{eff}$ at the minimum is then
\be 
S_0 
=\tau N_0\left(-\frac{M\Omega}{2}-\frac{e^2}{G} 
+\Omega e\right)\,. 
\label{S0} 
\ee 
 
Next we perform an expansion around the saddle point.  
We start by defining the fluctuation of the static $\rho$-field according to 
\begin{equation}
\rho = \overline{\rho}_0 + r 
=\overline{\rho}_0\left(1 + \frac{r}{\overline{\rho}_0}\right) 
\label{eq:34} 
\end{equation}
and by considering the generating functional (\ref{Zeff}), now written as
\be 
Z=\int_{-\overline\rho_0}^\infty \left[d r \right]  
\int_0^{\pi} \left[ d\theta \right] \exp (-S_{eff})\,.
\label{Zeff1} 
\ee 
Obviously this expansion is justified only if the quantum fluctuations in
(\ref{eq:34}) are small. We will show at the end of this 
Section that this is indeed the case.
 
To  proceed further we rewrite $S_{eff}$ in the form 
\be 
S_{eff} = \tau \sum_t G \left(\overline{\rho}_0 + r \right) + \mbox{Tr} \ln P 
-  \mbox{Tr} \ln \left[ 1\!\!1 + P \left (R_1 +R_2\right) \right]\ , 
\label{sefff}
\ee 
where 
\be 
R_1 = -q^-q^+ + ( \nabla_t^+ + e) ( \nabla_t^- - e) 
\ee 
and 
\be 
R_2 = G^2 r 
\,. 
\ee 
We set then  
\be 
S_{eff}=\sum_{r=0}^\infty S_r\,, 
\label{40}
\ee 
the term $S_0$ being the saddle point contribution, given by 
(\ref{S0}). This grows like $ \Omega^2$, but it contains  
also a term of order $\Omega$ and a term of order one, which should be kept
if an expansion in powers of $1/\Omega$ is sought for.
However, we prefer to stick to the definition 
(\ref{P-1}) for the operator $P$ and to compute the further contributions
to the expansion (\ref{40}) (the quantum fluctuations) by developing the last
logarithm in the right-hand-side of (\ref{sefff}): 
the terms thus obtained are naturally organized in powers of $M^{-1}$.
It is worth noticing that this expansion does not break 
the $U(1)$ invariance. 
 
Actually we shall confine ourselves to consider the
first and second order contributions in $\nabla_t\theta$ and $r$.  
 
The first order action stems from the term linear in $r$ and 
from the first term in the expansion of the logarithm and reads
\be 
S_1 = \tau G \sum_t r_t  
-  \mbox{Tr} \left[ P \left (R_1 +R_2\right ) \right]\,. 
\label{S1} 
\ee 
The explicit computation shows that all the terms linear in $r$ cancel out:  
hence the $r$-integration remains undefined.  
However the contributions arising from the second term in the expansion of the 
logarithm make the integral over $r$ well defined. The second order 
contribution will not be reported here and the details can be found in 
Ref.~\cite{Barbaro:2004nk}.
One is thus left with the result 
\be 
S_1=-\mbox{Tr} (PR_1) = \frac{1}{G} \left[ 1 + (M+\frac{3}{2} e)\tau \right]   
\tau \sum_{t=-\infty}^{\infty} \theta (-\nabla_t^+\nabla_t^-) \theta~.
\ee 
Note that these contributions are of order $\Omega$ and 1, namely are  
$O(1/\Omega)$ with respect to the saddle point one. 
 
In order to obtain
the Goldstone boson energies we must find out how they depend upon the
single particle energy $e$. 
For this purpose we have to perform in the integral expressing the  
generating functional $Z_1$ (associated with the action $S_1$) 
the $\theta$-integration, which is not
gaussian, because $\theta$ is compact. Yet we can  
choose $\nabla_t \theta$ as a new integration variable, thus rendering 
the integral gaussian, getting
\be 
-\frac{1}{N_0\tau}\ln Z_1= 
\frac{3}{4} e + \frac{M}{2}= 
\frac{3 e+G\Omega}{4}\,. 
\ee 

Finally, in order to deal with a specific system (e.g. a nucleus), 
the particle number must be fixed.
This can be accomplished by replacing $e$ with
\begin{equation}
\epsilon=e-\mu\ , 
\end{equation}
$\mu$ being the chemical potential and using
\begin{equation}
<\hat N>= \frac{1}{\tau N_0} \frac{\partial}{\partial \mu} \ln 
Z=  -\frac{1}{ \tau N_0} \frac{\partial}{\partial \epsilon} \ln Z\,, 
\label{numero} 
\end{equation}
where $\hat N$ is the particle number operator. Now,
replacing $<\hat N>$ with $2n$ and noticing 
that  $M$ does not depend on 
$\mu$~(since it is independent of $e$, see Eq. (\ref{Mmin})),  
Eq.~(\ref{numero}) becomes 
\begin{equation}
n = \frac{1}{N_0 \tau} \frac{\partial}{\partial \epsilon}  
(S_0-\ln Z_1)= 
-\frac{\epsilon}{G}+\frac{\Omega+3/4}{2}\,, 
\end{equation}
which gives  
\begin{equation}
\mu= G \left( n - \Omega/2-3/8\right) + e 
\label{mu1} 
\end{equation}
for the chemical potential. 
Hence, in the presence of the chemical potential, the energy of the system 
becomes 
\begin{eqnarray}
E_{n,0} &=& \frac{1}{\tau T}(S_0-\ln Z_1) + 2 \, \mu n  
\nonumber\\ 
&=&  2 e n - G n (\Omega - n+3/4)+\frac{G}{8}\left(5\Omega+\frac{9}{8}\right) 
\,,
\label{eq:E1mu} 
\end{eqnarray}
which reproduces the excitation spectrum of  
the pairing hamiltonian with good accuracy, the
relative difference between the exact ground state energy 
$-G[(\Omega+1)/2-e/G]^2$ (namely Eq.~\eqref{eq:E} with $s=0$ and $n=n_0$)
and the one corresponding to Eq.~(\ref{eq:E1mu})
being of order $1/\Omega$. 

Note that the same result can be obtained, rather than through the chemical 
potential, by 
introducing in the path integral the particle number projection operator 
\be 
{\cal P}_n=\int_{-\pi}^{+\pi} \frac{d\alpha}{2\pi} e^{-i (\hat N-2n)\alpha}\ ,
\label{Pn} 
\ee 
as shown in Ref.~\cite{Barbaro:2004nk}.

An important comment is in order on the validity of our expansion,
which depends upon the size of $\overline\rho_0$. 
We assess the latter by replacing in (\ref{ovrho0}) $e$ by $\epsilon$ 
and using (\ref{mu1})
(dropping the irrelevant term -3/8), thus obtaining
\be 
\overline\rho_0=n\left(\Omega-n\right)~,
\label{nOn} 
\ee 
which attains its maximum value $\overline\rho_0=\Omega^2/4$ for
$n=\Omega/2$.
This corresponds to the situation when the level where the pairs live is 
half-filled.
When this situation is almost realized, namely when the shell is 
neither fully filled nor almost empty, 
the functional integral defining $Z$ becomes gaussian and 
an expansion in $r/\overline\rho_0$ can be performed.
On the other hand $\overline\rho_0$ attains its lowest value when $n=1$
or $n=\Omega$. 
In this case there is no shift in (\ref{eq:34}) and the $\rho$ field 
acts only through its fluctuations, which are small, thus supporting the 
validity of the expansion.
It is then not surprising that our approach yields the (almost) 
correct excitation spectrum for any value of the pair number.

\subsection{The hamiltonian of the s-bosons}
\label{bosons}

We are now in a position of deriving the bosonic hamiltonian corresponding 
to the effective action previously obtained.
The most general, particle conserving, quartic hamiltonian 
for a system of $s$-bosons, confined to live in one single particle level,
reads in normal form (we omit for brevity the indices specifying the
single particle levels on which the $\hat b$ operators act)
\be
H(\hat{b}^{\dagger}, \hat{b}) =h_1 \hat{b}^{\dagger} \hat{b}  + 
h_2 \hat{b}^{\dagger}  \hat{b}^{\dagger}  \hat{b}\hat{b} ,
\label{Hbos}
\ee
$\hat{b}^{\dagger},\hat{b}$ being bosonic creation-annihilation operators 
acting in a Fock space and satisfying canonical commutation relations. 

The parameters $h_1,h_2$ should be fixed by requiring that the spectrum
of (\ref{Hbos}) coincides with the pairing Hamiltonian one. This
is indeed possible and it turns out that the choice 
\begin{equation}
\label{eq:h1h2}
h_1=2e-G\Omega~,\ \ \ \ \ h_2=G
\end{equation}
accomplishes the job.
However the Hamiltonian thus obtained, being intrinsically bosonic, 
patently violates the Pauli principle and therefore the condition $n< \Omega$
should be added {\it a posteriori}, when using (\ref{Hbos}) in dealing with a 
system of fermions. 

On the other hand in our framework this condition naturally emerges.
Indeed if we write the path integral associated to (\ref{Hbos})
\be
Z=\int [db^*db] \exp (-S) ,
\label{64}
\ee
where  
\be
S= \tau \sum_{t=1}^{N_0} 
\left[ b^*_{t+1} \nabla_t b_t + H(b_{t+1}^*,b_t) \right]~,
\ee
$b^*,b$ being holomorphic variables satisfying periodic boundary 
conditions in time, and perform the same saddle point expansion previously
illustrated, from the comparison with our effective action we get
\be
\label{eq:h}
h_1=2e-G\Omega-G/4 ,\,\,\,h_2=G ,
\,\,\,\,\,\,\,\,\,\,\,\,\,\,\,\,
\,\,\,\,\,\,\,\,\,\,\,\,\,\,\,\,
n<\Omega~.
\ee
The inequality in the above equation, expressing the Pauli principle, 
is necessary for the two path integrals \eqref{Zeff} and \eqref{64}
to coincide and follows from 
the positivity of $\bar\rho_0$ as given in Eq.~(\ref{nOn}): 
thus in our approach this condition, 
far from being artificial, is necessarily implied by the formalism itself.
Obviously the considerations following (\ref{40}) hold valid here as well, 
hence $h_1$ will be affected by an error of order $1/M$. 
Accordingly the values 
(\ref{eq:h}) essentially coincide with (\ref{eq:h1h2}), thus giving the
exact spectrum.

Summarizing, we have constructed an effective bosonic Hamiltonian, namely
\be
H(\hat{b}^{\dagger}, \hat{b}) =(2 e-G\Omega) \hat{b}^{\dagger} \hat{b}  + 
G \hat{b}^{\dagger}  \hat{b}^{\dagger}  \hat{b}\hat{b} ,
\label{Heff}
\ee
which yields the exact zero-seniority 
pairing spectrum and naturally incorporates, when treated within the path 
integral approach, the Pauli condition $n<\Omega$, keeping track of the 
fermionic nature of the composites.

\subsection{Seniority excitations: a matrix approach}
\label{sec:matrix}

In this Section we address the problem of the whole spectrum of the degenerate
pairing Hamiltonian, including 
the excited states with non zero seniority $v$. For convenience
we set $e=0$ and consider only the interaction Hamiltonian $H_P$.

The path integral approach followed in the $v=0$ case has not been pursued 
till now for $v\ne 0$ since it becomes quite cumbersome. 
Hence in this case it is preferable to work in the  hamiltonian formalism.

In the framework of the creation and annihilation operators the
commutator
\be
\left[ {\hat A}, ~{\hat A}^{\dag} \right]= 
\Omega\left(1- \frac{{\hat n}}{\Omega}\right)
\label{commutatoreA}
\ee
is non-canonical, thus rendering not trivial to find the eigenstates of $H_P$.
An approach circumventing this difficulty consists in 
employing again even Grassmann variables: in so doing the difficulty associated 
with the non-canonical nature of the commutator (\ref{commutatoreA}) 
disappears, since the associated Grassmann variables do commute.
 
Within this formalism we study
whether and how the Fock basis, set up to with determinants
of single particle states, can be reduced to a minimal dimension without
loosing the physical information we search for and 
whether this {\it minimal basis} can be expressed in terms of composite 
bosons~\cite{Barbaro:1999fh,Barbaro:1999mc}.

In seeking for the reduction of the basis dimensions, we shall be 
guided by the two main features of $H_P$, namely that it
is expressed solely in terms of the $\varphi$ and
is invariant for any permutation of the $\varphi$.

These properties, indeed, urge us to express also the vector of the basis 
in terms of the variables (\ref{eq:a}). The action of 
$H_P$ on these states is then obtained following 
the lines illustrated in \cite{Barbaro:1996pc}, the result being
\be
H_P \psi(\varphi^*) = \int [d\varphi'^* d\varphi'] K_P (\varphi^*,\varphi') 
e^{\sum{\varphi'^*\varphi'}}\psi(\varphi'^*) = E \psi(\varphi^*)~,
\label{eq_eigenvalue}
\ee
where the integration is over the even elements of the Grassmann algebra
and the kernel reads 
\be
K_P(\varphi^*,\varphi')= H_P(\varphi^*,\varphi') 
e^{\sum{\varphi^*\varphi'}}~.
\ee

We then attempt 
to diagonalize the $H_P$ associated with $n$ pairs in a basis 
set up with states represented as products of $n$ factors $\varphi^*$'s.
Since the number of these is 
${{\Omega}\choose{n}}$, the very large reduction
of the basis dimension entailed by the choice of the variables (\ref{eq:a}) is 
fully apparent: indeed in terms of fermionic degrees of freedom, 
the corresponding basis would have had a dimension ${{2\Omega}\choose{2n}}$.

Moreover, since the variables $\lambda$ anticommute,
each vector of this basis is antisymmetric in the exchange 
of any pair of fermions, thus fulfilling the Pauli principle.  

We now actually explore whether, for a given $\Omega$ and $n$, 
the eigenstates of $H_P$ can be cast into the form of a superposition of 
products of $n$ variables $\varphi^*$, namely 
\be
\psi (\varphi^*) = \sum_{m=1}
^{{\Omega}\choose{n}} 
\beta_m {[\varphi^*_{m_1}
\cdots \varphi^*_{m_n}]}_m~,
\label{psi_generale}
\ee
the index $m$ identifying the set of quantum numbers $\{ m_1, m_2\cdots m_n\}$ 
and the $\beta_m$ being complex coefficients.

From the condition (\ref{eq:s}) it follows that the 
number of distinct eigenvalues of $H_P$ is given by
$[\min \{n,\Omega-n\}+1]$,
each of them having a degeneracy
\cite{Ring}
\be
\delta_s
={{\Omega}\choose{s}} - {{\Omega}\choose{s-1}}~,
\label{degenerazione}
\ee
$s$ being, we recall, the number of broken pairs
(in our convention a binomial coefficient with a 
negative lower index vanishes).

We look for the eigenvalues of $H_P$ by 
diagonalizing the symmetric matrix
${\cal H}_P\equiv (E-H_P)/G$, hence dimensionless, given by
\be
\left(\begin{array}{ccc}
{\cal E} +n &  & 0\lor 1\\
&\ddots  & \\
0\lor 1 & & {\cal E} +n  
\end{array}
\right) \ .
\label{matrice_generica}
\ee
The dimension of the above is ${{\Omega}\choose{n}}$
and the symbol $0\lor 1$ indicates that the upper (lower)
triangle of the matrix is filled with zeros and ones.  
Indeed the matrix elements of ${\cal H}_P$ are one when the 
bra and the ket differ by the quantum state of one 
(out of $n$) pair, otherwise they vanish. 
The diagonal matrix elements simply count the number of pairs and
${\cal E}=E/G$.

We now look for a further dimensional reduction of the basis such that the
resulting matrix  Hamiltonian has the same eigenvalues of the original one, 
however all being non degenerate. For this purpose
an elementary combinatorial analysis shows that the number of ones 
in each row 
(column) of the matrix (\ref{matrice_generica}) is given by $n(\Omega -n)$.
Indeed a non-vanishing matrix element has the row specified by $n$ 
indices whereas, of the indices identifying the column, $n-1$ should be 
extracted from those fixing the row in all the possible ways, which
amounts to $n$ possibilities. The missing index should then be selected 
among the remaining $\Omega-n$ ones: hence the formula $n(\Omega - n)$ 
follows.  

Note that the cases with $n$ and $\Omega-n$ pairs are equivalent,
an occurrence which is also manifest in the exact
spectrum \eqref{eq:E}.
Hence in the following we shall confine ourselves 
to consider $n\le \frac{\Omega}{2}$ only.

To write down explicitly the matrix (\ref{matrice_generica}) it
is convenient to divide the set of the $\Omega$ even Grassmann variables, 
whose quantum numbers 
identify the levels where the $n$ pairs are placed, into two 
subsets: one with $\Omega - n$ and the other with $n$ elements 
(to be referred to as I and II, respectively).
This partition leads to a 
basis with dimension $n+1$, independently of the value of $\Omega$.
The associated eigenvalues correspond to the breaking of 0, 
1, 2 $\cdots n$ pairs. 

Indeed in this instance the ${{\Omega}\choose{n}}$ entries of 
each row and column of the matrix can then
be grouped into $n+1$ sets, the first one corresponding to the $n$ pairs placed
in the $\Omega-n$ levels of I, the remaining $n$ levels of II being empty
(see Fig.~1a).

\begin{figure}
\epsfig{figure=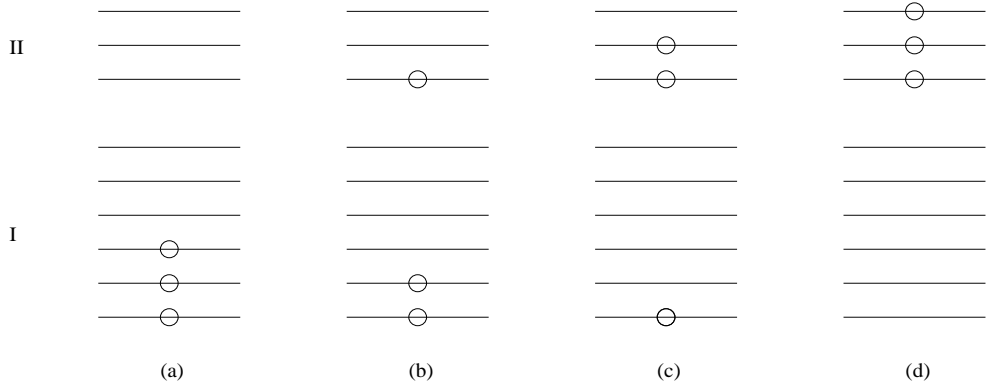,width=13cm,height=5cm}
\caption[ ]{The figure shows, in the specific case $n$=3 and $\Omega$=9,
the partition of the $\Omega$ levels in the set I (with $\Omega -n$=6 levels)
and II (with $n=3$ levels).}
\end{figure}
 
The number of configurations belonging to this first set is 
$d_0={{\Omega-n}\choose{n}}$. 
To the second set are associated configurations with $n-1$ pairs in I and 
one pair in II (see Fig.~1b), their number being 
$d_1=n {{\Omega-n}\choose{n-1}}$. In general the $(k+1)$-th set embodies 
configurations with $n-k$ pairs in I and $k$ pairs in II, their number being

\be
d_k = {{n}\choose{k}} {{\Omega-n}\choose{n-k}} \ \ \ \mbox{with}\ 
0\leq k\leq n\ \ (k\ \mbox{integer}).
\label{dk}
\ee
Clearly the total number of configurations is
\be
\sum_{k=0}^n d_k = {{\Omega}\choose{n}} \ .
\ee

With this organization of the levels 
the matrix (\ref{matrice_generica}) splits into $(n+1)^2$ rectangular blocks 
$B_{kj}$ (with $0\leq k,j\leq n$) of dimension $d_k\times d_j$ and,
since the pairing Hamiltonian only connects states differing by the 
quantum number of one pair, the blocks with $|k-j|\geq 2$ will have 
vanishing elements: the matrix thus becomes {\it block-tridiagonal}. 

This suggests the 
introduction of the following orthonormal set
of $n+1$ even commuting variables (the composites)

\be
\Phi_k^* = \frac{1}{\sqrt{d_k}} \sum_{m=1+s_{k-1}}^{s_k} 
\left[\varphi^*_{m_1}...\varphi^*_{m_n}\right]_m \ ,
\label{Phik}
\ee
where $k$ varies as in (\ref{dk}), $s_j = \sum_{l=0}^j d_l$, 
being $s_{-1}=0$ and $m$ again identifies the set $\{m_1, m_2 \cdots m_n\}$. 
Note that the variables (\ref{Phik})
have in general an index of nilpotency higher than one. 

The definition (\ref{Phik}) also reflects our desire that 
the composite bosons keep as much as possible of the 
symmetry of $H_P$. And indeed the $\Phi^*_k$, while not fully 
symmetric with respect to the interchange of the $\varphi^*$, turn out 
to be invariant with respect to the 
interchange of the $\varphi^*$ belonging either to the set I or to the set II. 
This symmetry property  
enforces the maximum coherence among the components of $\Phi^*_k$. 
It is remarkable that 
composite variables corresponding to combinations of the 
$\varphi^*_{m_1} \varphi^*_{m_2} \cdots \varphi^*_{m_n}$ different from 
\eqref{Phik} not only 
hold a lower symmetry than the one displayed by (\ref{Phik}), but may also 
lead, as we have verified in some instances, to the wrong eigenvalues. 

Now in the minimal basis (\ref{Phik}) ${\cal H}_P$ is 
represented by a $(n+1)\times(n+1)$ matrix whose generic element 
$\left(M_n\right)_{ki}\equiv <\Phi_k^*|{\cal H}_P|\Phi_i^*>$ 
obtains by summing the elements of the
block $B_{ki}$ of the ${\Omega\choose n}\times{\Omega\choose n}$ matrix,
but for the normalization factor $1/\sqrt{d_k d_i}$.
The sum is performed by recognizing that all the blocks 
have the same number of ones in each row.
Specifically, in the diagonal block $B_{kk}$ the number of ``ones'' 
in each row is $(n-k)(\Omega-2n+2k)$. In the upper diagonal block
$B_{k,k+1}$ each row contains instead
\be
c_k=(n-k)^2
\label{ck}
\ee
ones. Since the total number of ones in each row of the matrix 
(\ref{matrice_generica}) is $n(\Omega-n)$, the number of ones in the 
rows of the lower diagonal block $B_{k,k-1}$ will be 
\be
b_k=k(\Omega-2n+k)\ . 
\label{bk}
\ee
As a consequence the non vanishing elements of the matrix $M_n$ turn out to be

\bea
\left(M_n\right)_{k,k+1} &=& <\Phi^*_k|{\cal H}_P|\Phi^*_{k+1}> = 
\sqrt{\frac{d_k}{d_{k+1}}} c_k~,
\\
\left(M_n\right)_{k+1,k} &=& <\Phi^*_{k+1}|{\cal H}_P|\Phi^*_k> =
\sqrt{\frac{d_{k+1}}{d_k}} b_{k+1}
\eea
and
\be
\left(M_n\right)_{kk} = <\Phi^*_k|{\cal H}_P|\Phi^*_k> = a_k \equiv 
{\cal E} + n + n(\Omega-n) - b_k - c_k\ .
\ee
Clearly $\left(M_n\right)_{k,k+1}=\left(M_n\right)_{k+1,k}$, 
since the operator ${\cal H}_P$ is Hermitian
and the basis (\ref{Phik}) is orthonormal.
The matrix thus becomes {\it tridiagonal}, reading

\bea
M_n = \left(
\begin{array}{ccccccc}
a_0 & \sqrt{\frac{d_0}{d_1}} c_0 &        0     & \cdot   & \cdot  & \cdot 
& \cdot\\
\sqrt{\frac{d_0}{d_1}} c_0 &  a_1 & \sqrt{\frac{d_1}{d_2}} c_1  &    0    
& \cdot  & \cdot & \cdot\\
   0    & \sqrt{\frac{d_1}{d_2}} c_1 &  a_2  & \sqrt{\frac{d_2}{d_3}} c_2 
&    0   & \cdot & \cdot\\
 \cdot  & \cdot   &     \cdot    & \cdot   & \cdot  & \cdot &\cdot\\   
 \cdot  &  0      &\sqrt{\frac{d_{k-1}}{d_k}}c_{k-1}& a_k 
&\sqrt{\frac{d_k}{d_{k+1}}} c_k & 0     & \cdot\\   
 \cdot  & \cdot   & \cdot        & \cdot   & \cdot  & \cdot &\cdot\\   
 \cdot  & \cdot   & \cdot        & \cdot   & 0      
&\sqrt{\frac{d_{n-1}}{d_n}}  c_{n-1} & a_n   
\end{array} \right)
\nonumber
\eea
and the associated eigenfunctions should be expanded in terms of 
the composite variables (\ref{Phik}), namely
\be
\psi (\Phi^*) = \sum_{k=0}^n u_k \Phi^*_k \ .
\label{psi_Phi}
\ee

It is convenient to cast the coefficients of the expansion 
(\ref{psi_Phi}), fixed by the eigenvalue equation

\be
M_n \vec u = 0 \ ,
\label{eqwk}
\ee
into the form
\be
u_k \equiv \sqrt{d_k} w_k \ .
\ee
Thus the ${\Omega\choose n}$ equations for the $\beta$'s reduce
to $n+1$ equations for the $w$'s and  
the associated eigenvalues ${\cal E}$ obey the secular equation
\be
D_n = \det \left(M_n\right) = 0 \ ,
\label{DetD0}
\ee
where ${\cal E}$ enters into the diagonal matrix elements $a_k$.
 
Now from the general theory of symmetric tridiagonal matrices
one knows that Eq.~(\ref{DetD0}) has $n+1$ 
{\em distinct} and {\em real} 
roots and these are found by applying the recursive relation 

\be
D_n = a_n D_{n-1} - \frac{d_{n-1}}{d_n} (c_{n-1})^2 D_{n-2}
\label{recD}
\ee
for increasing values of $n$. Hence we have

\bea
D_0 &=& y
\nonumber\\
D_1 &=& y(y-\Omega)
\nonumber\\
D_2 &=& y(y-\Omega)[y-2(\Omega-1)]
\nonumber\\ 
\cdots
\nonumber\\
D_n &=& y(y-\Omega)[y-2(\Omega-1)][y-3(\Omega-2)] ... [y-n(\Omega-n+1)] 
\label{D_n}\ ,
\eea
where
\bea
y &=& n(\Omega-n)+{\cal E}+n \ .
\eea
We thus see that $D_n$ has all the zeros of $D_{n-1}$ plus an extra one for
$y=n(\Omega-n+1)$.

Moreover (\ref{D_n}) allows us to write down for the general solution of 
(\ref{DetD0}) the expression 
\be
y=p(\Omega-p+1) \ \ \ \ \ \ \ \mbox{with} \ 0\le p \le n ~.
\ee
Hence {\it the well-known formula
for the spectrum of the pairing Hamiltonian} 

\be
{\cal E} = -(n-p)(\Omega-n-p+1) ~,
\ee
{\it is recovered},
the index $p$ coinciding with the pair seniority quantum number $s$.

Let us now consider the eigenfunctions of $\hat H_P$.
For a tridiagonal matrix, a recursive relation among the eigenvectors 
components, similar to (\ref{recD}), can also be established.
In the specific case of the matrix $M_n$ it reads
\be
d_k c_k w_{k+1} = - d_{k-1} c_{k-1} w_{k-1} - d_k a_k w_k \ ,
\label{recw}
\ee
where again $0 \le k \le n$ and quantities with negative indices 
are meant to be zero. 

Thus for the lowest eigenvalue $y=0$ ($p=0$, zero seniority) we have

\be
w_0^{(s=0)}=w_1^{(s=0)}=....=w_n^{(s=0)} \ ,
\label{w0}
\ee
namely the collective state
\be
\psi_{s=0} ={\Omega\choose n}^{-1/2} \sum_{k=0}^n \sqrt{d_k} \Phi^*_k
= {\Omega\choose n}^{-1/2} \sum_{m=1}^{\Omega\choose n} 
\left[\varphi^*_{m_1}...\varphi^*_{m_n}\right]_m \ .
\label{sen0}
\ee
Indeed in (\ref{sen0})
all the components of the wave-function, i.e. the monomials
$\left[\varphi^*_{m_1}...\varphi^*_{m_n}\right]_m$, are coherently summed up.
This state obtains for a specific partition of the levels 
defining the matrix $M_n$. However, any other partition would lead to
the same result, being all the weights of the components equal. As a 
consequence the state (\ref{sen0}) is non degenerate.

Concerning the second eigenvalue $y=\Omega$ ($p=1$, pair seniority $s=1$),
according to Eq.~(\ref{recw}) the components of its eigenstate are
\be
w_k^{(s=1)} = {\cal N}_1 \left(k\Omega - n^2\right) \ .
\label{w2}
\ee
Finally the components of the state associated with a generic pair
seniority $s$ turn out to read
\bea
w_k^{(s)}& = & {\cal N}_s \sum_{j=0}^{s} (-1)^j \frac{ {k \choose j} 
{s \choose j} 
{\Omega -s +1 \choose j}}{{n \choose j}^2}=\nonumber\\
&=& {\cal N}_s \ _3F_{2} \left( -k, -\Omega +s -1, -s; ~
-n, -n ;~ 1\right) 
\label{wk}
\eea
$ _3F_{2}$ being a generalized hypergeometric function.
In the above the binomial $k \choose j$ is meant to vanish when 
$j > k$. 

In particular for the vector of the maximum seniority $\vec w^{(s=n)}$, 
corresponding to $y=n(\Omega-n+1)$ ($s=n$), 
one has 
\be
w_k^{(s=n)} = {\cal N}_{n} \left[(-1)^k 
\frac{(\Omega-2n+k)!(n-k)!}{(\Omega-2n)!}\right] \ .
\label{wn}
\ee

In (\ref{w2}), (\ref{wk}) and (\ref{wn}) ${\cal N}_1$, ${\cal N}_s$ and 
${\cal N}_{n}$ are normalization constants. 

It is interesting to reobtain in the present formalism the following important 
result \cite{BMII}: 
for values of $\Omega$ large with respect to $n$, 
the number of dominant components in
$\vec w^{(s)}$ decreases with $s$, reflecting the weakening
of collectivity with increasing seniority.
Indeed in the limit $\Omega>>n$ the eigenvalues are

\be
y\simeq s\Omega 
\label{yv}
\ee
and moreover 
\be
c_k << \Omega \ ,
\ \ \ 
\frac{d_{k-1}}{d_k} c_{k-1} \simeq k\Omega
\ \ \ 
\mbox{and}
\ \ \ 
a_k \simeq \left(s-k\right)\Omega \ ,
\ee
being $0\leq s\leq n$.

By inserting the above limits in (\ref{recw}) the $w_k^{(s)}$ 
components corresponding to the eigenvalues (\ref{yv}) are found to be 
\be
w_0^{(s)} = w_1^{(s)} = ... = w_{s-1}^{(s)} = 0
\ee
and
\be
w_k^{(s)} = \frac{{k\choose s}}{{n\choose s}} w_n^{(s)}
\ ,\ \ \ \ \ \ \ k=s,...,n \ .
\label{wkj}
\ee

We thus see that the state with seniority 
$s$ has indeed, in the basis of the $\sqrt{d_k}\Phi^*_k$ and
in the large $\Omega$ limit, $s$ vanishing components,
the remaining $n-s+1$ ones being expressed, through (\ref{wkj}),
via the single component $w_n^{(s)}$.
In other words, when $\Omega$ is large, the collectivity of a state 
with seniority 
$s$ decreases as $s$ increases, because its components become 
fewer and fewer and, furthermore, 
the surviving components are more and more expressible through a single one.

\section{The non degenerate case}
\label{sec:nondeg}

We now turn to explore the case of the pairing Hamiltonian $\hat H$ acting 
on a set of non degenerate single particle levels.
The actual situation one faces in applying the pairing Hamiltonian to
real systems corresponds indeed to having more than one energy $e_\nu$
in \eqref{eq:HP}. In this respect nuclei and metals represent two extreme
situations of the non-degenerate case: in the former
a major shell is typically split into five or 
six single particle levels of different angular momenta, in the latter
the number of non-degenerate levels entering into a band corresponds to
a significant fraction of the Avogadro number. Moreover
in a heavy nucleus the number of pairs living in a level may be as large as,
say, eight while in a metal is one. Recently,
a renewed and widespread interest for the pairing problem
in the non-degenerate frame has flourished in connection with 
the physics of ultrasmall metallic grains, possibly 
superconducting~\cite{Sierra:1999rc},
and of Bose-Einstein condensation~\cite{Dukelsky:2001fe}.

As we shall illustrate,
in the non-degenerate case one more type of excitations occurs with respect to the
degenerate one, corresponding to promoting pairs above the Fermi sea: 
these are zero seniority excitations which, in the BCS language, are described 
as four quasi-particle states (more precisely, as two quasi-particles and two 
quasi-holes states).
These are conveniently classified in terms of a new quantum number,
the ``like-seniority'', 
and reflect the occurrence of a quantum phase
transition.

Furthermore, and more importantly, in the non-degenerate case critical values $G_{\rm cr}$
of the coupling arise which split the physics into a region governed by the mean field
(when $G<G_{\rm cr}$) and a region governed by the pairing force (when $G>G_{\rm cr}$).
This competition does not show up in the degenerate case.

\subsection{The Richardson exact equations}

In this Section we derive in an alternative way the well-known Richardson 
equations~\cite{Rich}, using the Hamiltonian formalism expressed in the 
framework of even Grassmann variables previously introduced.
As already mentioned, this is particularly useful for avoiding the non 
canonical commutators, which would naturally appear in the standard Hamiltonian
treatment of the pairing problem.
 
In the following we consider the simpler case of zero-seniority states, although the
approach can be generalized to include seniority excitations by introducing other
composite variables, as done for one pair in Ref.~\cite{Barbaro:2001ir}.

Starting from the normal kernel of the Hamiltonian (\ref{eq:HP})
 written in terms of Grassmann variables
\begin{equation}
  \label{y7}
  H= \sum_{\nu=1}^L e_\nu \sum_{m_\nu=-j_\nu}^{j_\nu}
\lambda^*_{j_\nu m_\nu} \lambda_{j_\nu m_\nu} -
G \sum_{\mu,\nu=1}^L \Phi^*_\mu\Phi_\nu \ ,
\end{equation}
we search for eigenstates of $n$ pairs of fermions 
in the $s$-quasibosons subspace as products of $n$ factors, namely
\begin{equation}
  \label{eq:1}
\psi_n(\Phi^*)(m) = \prod_{k=1}^n {\cal B}^*_k(m) \ ,
\end{equation}
where
\begin{equation}
  {\cal B}^*_k(m) = \sum_{\nu=1}^L \beta_\nu^{(k)}(m) \Phi^*_\nu 
\label{psin}
\end{equation}
is a superposition of $s$-quasibosons placed in all the available levels
and the index $m=(m_1,\dots,m_n)$ labels the 
unperturbed configuration from where the state develops as the pairing force
is switched on. The set of values of $m$ corresponds to the possible states
available to the system.
When no confusion arises the index $(m)$ will be dropped.

It is convenient to start from the effective Hamiltonian
\begin{equation}
  \label{eq:y8}
  {\cal H}_{\rm eff}(\varphi^*,\varphi) =
\sum_{\nu=1}^L 2 e_\nu \sum_{m_\nu=1/2}^{j_\nu}
\varphi^*_{j_\nu m_\nu} \varphi_{j_\nu m_\nu} -
G \sum_{\mu,\nu=1}^L \Phi^*_\mu\Phi_\nu \ ,
\end{equation}
coincident with (\ref{y7}) in the $s$-quasibosons subspace spanned by 
the states (\ref{eq:1}).
Indeed while terms like $\lambda^*\lambda$ count the number of particles, 
$\varphi^*\varphi\equiv\lambda^*\lambda^*\lambda\lambda$ counts the number 
of pairs.
The eigenvalue equation can then be written as
\begin{equation}
  \label{eq:13}
\int [d{\varphi'^*} d\varphi']{\cal H}_{\rm eff}
(\varphi^*,\varphi')
\exp\left(\sum_{\nu,m_\nu}
(\varphi^*_{j_\nu m_\nu}+\varphi'^*_{j_\nu m_\nu})\varphi'_{j_\nu m_\nu}\right)
\psi_n(\Phi'^*) = E_n  \psi_n(\Phi^*)\ ,
\end{equation}
since in the expansion of the exponentials 
only the even powers, hence only
the $\varphi$ variables, survive.
By performing the integrals over the $\varphi'$s
one gets then
\begin{equation}
E_n(m) = \sum_{k=1}^n E_k(m) \ ,\ \ \ \
\beta_\mu^{(k)}(m) = \frac{C_k(m)}{2 e_\mu-E_k(m)} \ ,
\label{eq:beta}
\end{equation}
where $C_k(m)$ are normalization factors (see \cite{Barbaro:2003bb} for their
expression) and $E_k(m)$ (henceforth referred to as {\em ``pair energies''})
are 
the solutions of the non-linear system
\begin{equation}
\sum_{\mu=1}^L \frac{\Omega_\mu}{2 e_\mu-E_k}-
\sum_{l=1,l\neq k}^n \frac{1}{E_l-E_k}=\frac{1}{G} \ ,
\ \ \ \ \ k=1,\cdots n\ .
\label{eq:system}
\end{equation}
We thus see that the Grassmann variables  formalism enables us to recover
the same equations obtained within, for instance, the quasi-spin framework
and usually known as Richardson's equation.


Since the equations (\ref{eq:system}) deal with pairs of fermions 
(e.g. nucleons)
coupled to an angular momentum $J=0$, their eigenvalues, given
by $E=\sum_k E_k$, are those of the  
zero-seniority states ($v=0$). Importantly, these eigenvalues
display different degrees of collectivity: hence they are
conveniently classified in terms of the latter.

To clarify this point we introduce a number $v_l$,  
that counts in a given state the number of particles prevented
to take part into the collectivity, not because they are blind to the
pairing interaction (indeed they are coupled to $J=0$), 
but because they remain trapped in between the single particle levels, 
even in the strong coupling regime.
This number is directly linked to the number $N_G$, first introduced by
Gaudin in Ref.~\cite{Gau95}, through the relation $v_l=2 N_G$ and
might be considered as a sort of ``like-seniority'' (hence the notation
$v_l$)~\footnote{Note however that, whereas the seniority $v$ counts the
fermions coupled to $J\neq 0$, the number $v_l$ refers to $J=0$ pairs.} 
since it reduces to the standard seniority $v$ for large $G$: as 
pointed out in \cite{Roman}, $N_G$ has the significance of the number of pair 
energies which remain finite as $G$ goes to infinity.

Specifically we shall ascribe the value $v_l$= 0 to the fully collective state,
$v_l$=2 to a state set up with a trapped pair energy while the others
display a collective behavior, $v_l$=4 to the state with
two trapped pair energies and so on.

Worth noticing is that in the degenerate case $e_\mu=e$ the result 
(\ref{eq:E}) for
vanishing seniority is easily recovered by multiplying Eq.~(\ref{eq:system}) by
$(2 e-E_k)$ and summing over $k$. Indeed, exploiting the identity
\begin{equation}
\sum_{k=1}^n \sum_{l(\neq k)=1}^n \frac{E_k}{E_l-E_k}= -\frac{1}{2}n(n-1)~,
\end{equation}
one immediately gets
\begin{equation}
E= 2 e n- G  n \left(\sum_{\mu=1}^L \Omega_\mu - n+1\right)~,
\end{equation}
which coincides with (\ref{eq:E}) for $s=0$.

In the non degenerate case an exact analytic solution of the Richardson system 
cannot be obtained and one has to resort to numerical calculations.
However, approximate analytic expressions for both the ground and the excited 
states can be
obtained and they help in understanding the occurrence of critical 
phenomena. This will be illustrated in the next two Sections for the cases
of 1 and 2 pairs.

\subsection{One pair}

We start by addressing the problem of one pair living in many single 
particle levels and show that, even in this case, 
a critical value of the coupling constant $G$ can occur which separates 
two different regimes~\cite{Barbaro:2002fi}.

For $n=1$ the Richardson system (\ref{eq:system}) reduces to the equation
\begin{equation}
  \label{eq:20}
   \sum_{\nu=1}^L \frac{\Omega_\nu}{2 e_\nu - E} = \frac{1}{G}\ ,
\end{equation}
which yields $L$ eigenvalues $E(m)$ ($1\leq m\leq L$), 
while the wave function
\begin{equation}
  \label{eq:14}
  \psi(\Phi^*)(m) = \sum_{\nu=1}^L \beta_\nu(m) \Phi^*_\nu
\end{equation}
is characterized by the coefficients $\beta_\nu(m)=C(m)/(2 e_\nu-E(m))$,
according to Eq.~(\ref{eq:beta}).

It is straightforward to solve equation (\ref{eq:20}) numerically: 
the solutions can be graphically displayed as the intersections
of the left-hand-side of (\ref{eq:20}) with the straight line $E=1/G$~\cite{Rowe}.
Two classes of states 
appear: the first one embodies the lowest energy state, 
which lies below the lowest single particle level for an attractive 
interaction and represents a collective state when $G$ is large with respect
to the single particle energies;
the other contains the so-called ``trapped'' solutions, 
which lie in between the single particle levels.

Concerning the wave function, the one corresponding to the lowest 
eigenvalue, if $G$ is large enough to organize a collective motion,
has all the $\beta$ coefficients sizable,
reflecting the high degree of collectivity of the state.
On the contrary, in the eigenstates of the trapped solutions only one
$\beta$-coefficient dominates, so that their wave functions are close to the
ones of an unperturbed state.

Let us now consider the collective state in the strong coupling regime.
Here it is convenient to recast Eq.~(\ref{eq:20}) as 
\begin{equation}
\label{eq:6.1bis}
\sum_\nu\frac{\Omega_\nu}{2e_\nu-E} =
-\frac{1}{E-2\bar e} \sum_\nu \frac{\Omega_\nu}{1-2
\dfrac{e_\nu-\bar e}{E-2\bar e}}=\frac{1}{G}
\end{equation}
and expand in the
parameter $2(e_\nu-\bar e)/(E-2\bar e)$.
Defining
\begin{equation}
\bar e=\frac{\sum_\nu \Omega_\nu e_\nu}{\Omega}
\label{ebar}
\end{equation}
in leading order we get
\begin{equation}
E_0=2\bar e-\Omega G\ ,\ \ \ \ \ \Omega=\sum_\nu\Omega_\nu~,
\end{equation}
which coincides with the degenerate case value.
Owing to the definition (\ref{ebar}), the next-to-leading order 
correction vanishes.
To proceed further we rewrite (\ref{eq:6.1bis}) as 
\begin{equation}
  \label{eq:n6.3}
  -\frac{\Omega}{E-2\bar e}\sum_{n=0}^\infty
  \frac{2^n M^{(n)}}{(E-2\bar e)^n}=
-\frac{\Omega}{E-2\bar e}\sum_{n=0}^\infty
\left(\frac{G\Omega}{E-2\bar e}\right)^n m^{(n)} \alpha^n=
\frac{1}{G}\;,
\end{equation}
where the generalized moments of the distribution of
single particle levels $e_\nu$, namely
\begin{equation}
  \label{eq:n6.2}
M^{(n)}=\sigma^n m^{(n)}=\frac{1}{\Omega}\sum_\nu\Omega_\nu(e_\nu-\bar e)^n\ ,
\end{equation}
have been introduced together with the dimensionless expansion parameter
\begin{equation}
  \label{eq:n6.7}
  \alpha=\frac{2\sigma}{G\Omega}
\end{equation}
and the variance
\begin{equation}
  \label{eq:n6.2a}
\sigma =\sqrt{\frac{1}{\Omega}\sum_\nu\Omega_\nu(e_\nu-\bar e)^2}~.
\end{equation}
The strong coupling regime then corresponds to $\alpha\ll 1$.
Note that the second moment is the square of the variance $\sigma$, 
whereas the third, $M^{(3)}=\sigma^3\gamma$, and fourth,
$M^{(4)}= (c+3) \sigma^4$, moments are related to the skewness $\gamma$
and to the kurtosis $c$ of the distribution, respectively.
Note also that $\alpha=\sigma M^{-1}$, being $M^{-1}$ the dimensionful
expansion parameter introduced in Eq.~(\ref{Mmin}).

In a perturbative frame, setting $E^{(n)} = E^{(n-1)} +\delta$
and linearizing in $\delta$, we get for the lowest energy
\begin{equation}
  \label{eq:n6.18}
  E = 2\bar e - G\Omega \left[1+\alpha^2-\gamma\alpha^3+(1+c)\alpha^4 +
{\cal O}(\alpha^5)\right]\ ,
\end{equation}
an expression valid for $\alpha\ll 1$ and for any single particle 
energy distribution, but in particular when the fluctuation of the latter 
around $\bar e$ is small.

The evaluation of the trapped eigenvalues can be easily performed 
numerically. This approach however hides some interesting properties
of the solutions, which correspond to a transition from the mean field
to the pairing domain, as we shall illustrate in the following.

The number of trapped solutions of the secular equation (\ref{eq:20}) 
for one pair
and $L$ single particle levels is $L-1$. They can be identified by a quantum
number $\nu=2,\dots L$, which labels the level where the pair
lives when $G=0$ ($\nu=1$ corresponds to the lowest unbounded energy).
Since they are trapped, namely
\begin{equation}
 2e_{\nu-1}<E^{(\nu)}<2e_\nu~,
\end{equation}
a new variable $z^{(\nu)}\in(0,1)$ can be defined according to
\begin{equation}
  \label{eq:4.01}
  E^{(\nu)}=2e_{\nu-1}+2 z^{(\nu)}(e_\nu-e_{\nu-1})~.
\end{equation}

If we now isolate in the secular equation the terms associated with the poles in
$2e_{\nu-1}$ and $2e_\nu$, we get
\begin{equation}
  \label{eq:4.02}
  \frac{\Omega_{\nu-1}}{z^{(\nu)}}+\frac{\Omega_\nu}{z^{(\nu)}-1}=
\sum_{\mu(\not=\nu,\nu-1)=1}^L
  \frac{\Omega_\mu}{\dfrac{e_\mu-e_{\nu-1}}{e_\nu
      -e_{\nu-1}}-z^{(\nu)}}-\frac{2(e_\nu
      -e_{\nu-1})}{G}~.
\end{equation}

Next we let the discrete variable $\nu$ become continuous
by means of the Euler-McLaurin formula~\cite{SaGe-60-B}, which 
replaces the sum with an integral, i.e.
\begin{equation}
\sum_{\nu=a}^{b} f(\nu) \simeq \frac{1}{2} f(a) + \frac{1}{2} f(b) +
\int_{a}^b f(u) du \ .
\label{maclaurin}
\end{equation}
The formula (\ref{maclaurin}) holds valid if $f(\nu)$, $\nu$ to be viewed as
a complex variable, is analytic in the strip $a\leq \mbox{Re}\nu \leq b$ ($a$ 
and $b$ being integers).

To proceed further one must specify the single particle energies and the
associated degeneracies.
Here we shall first consider the case of a harmonic oscillator well, and
later generalize the results.
The harmonic oscillator unperturbed energies are
labeled by an index $k=0,1,\cdots {\cal N}-1$ and read
\begin{equation}
e_k\equiv \tilde e_k \hbar \omega_0 = (k+ 3/2) \hbar \omega_0\ ,\ \ \ \ \ \ \ 
(k=0,1,\cdots {\cal N}-1)\label{ekHO}
\end{equation}
with associated pair degeneracies
\begin{equation}
\Omega_k = (k+1)(k+2)/2\ .
\label{OmkHO}
\end{equation}
The total degeneracy $\Omega$, the average energy $\bar e$ and the variance 
$\sigma$ are found to be
\begin{equation}
  \label{eq:3.4.4}
  \Omega=\frac{1}{6}{\cal N}({\cal N}+1)({\cal N}+2)\ ,
\end{equation}
\begin{equation}
  \label{eq:3.4.3}
  \overline e=\frac{3}{4}({\cal N}+1)\hbar \omega_0
\end{equation}
and
\begin{equation}
\sigma^2 = \frac{3}{80}({\cal N}-1)({\cal N}+3) (\hbar \omega_0)^2\ ,
\end{equation}
respectively, entailing the following expression for the expansion parameter:
\begin{equation}
\alpha^{(h.o.)}=
\frac{3\hbar\omega_0}{G} \sqrt{\frac{3}{5}}
\frac{\sqrt{({\cal N}-1)({\cal N}+3)}}{{\cal N}({\cal N}+1)({\cal N}+2)}\ .
\label{eq:alpha_ho}
\end{equation}
Using the dimensionless quantities $\tilde G=G/\hbar\omega_0$ and
$\tilde z=z/\hbar\omega_0$ we then rewrite (\ref{eq:4.02}) 
in the Euler-McLaurin approximation obtaining ($k$ labels the single particle 
levels)
\begin{equation}
  \label{eq:4.020}
  \frac{\Omega(k-1)}{\tilde z{(k)}}+\frac{\Omega(k)}{\tilde z(k)-1}=
  \varphi^{\rm E-McL}(k,\tilde z{(k)})
\end{equation}
with
\begin{eqnarray}
  \varphi^{\rm E-McL}(k,\tilde z)&=&
  -\frac{2}{\tilde G}-\frac{3}{4}(4k+2\tilde z+3)+\frac{1}{4}
  ({\cal N}-1)({\cal N}+2k+2\tilde z+3)
\nonumber\\
  &+&\frac{k(1-k)}{4(\tilde z+1)}-\frac{(k+2)(k+3)}{4(\tilde z-2)}\nonumber\\
  &-&\frac{1}{2(k+\tilde z-1)}-\frac{{\cal N}({\cal N}+1)}
{4(k+\tilde z-{\cal N})}\nonumber\\
  &+&\frac{1}{2}(k+\tilde z)(k+\tilde z+1)\log\left|\frac{(\tilde z+1)
(k+\tilde z-{\cal N})}{(\tilde z-2)(k+\tilde z-1)}
    \right|\ .
  \label{eq:4.021}
\end{eqnarray}
The Euler-McLaurin approximation provides the key for studying analytically
the limiting case ${\cal N}\to \infty$, which is
helpful in shedding light on the properties of the trapped eigenvalues. 
For this purpose we define
\begin{equation}
  \label{eq:9.201}
  \lambda=\frac{k}{{\cal N}}~,
\end{equation}
which, when ${\cal N}$ is large, tends to become a continuous variable
in the interval $(0,1)$.

Moreover we express the coupling constant $\tilde G$ in terms of
$\alpha$ according to (\ref{eq:alpha_ho}).
Then Eq.~(\ref{eq:4.020}) becomes
\begin{equation}
\phi(\tilde z)=\xi_\infty(\lambda)
  \label{eq:4.030}
\end{equation}
with
\begin{equation}
 \phi(\tilde z)\equiv
 \frac{1}{\tilde z}+\frac{1}{\tilde z-1}+\frac{1}{2(\tilde z+1)}+
\frac{1}{2(\tilde z-2)}
  -\log\left|\frac{\tilde z+1}{\tilde z-2}\right|
\end{equation}
and
\begin{equation}
\xi_\infty(\lambda) \equiv \frac{1}{\lambda}+\log\frac{1-\lambda}{\lambda}+
  \frac{1-\frac{8}{3}\sqrt{\frac{5}{3}}\alpha}{2\lambda^2}\;.
\end{equation}

\begin{figure}[ht]
  \begin{center}
    \epsfig{file=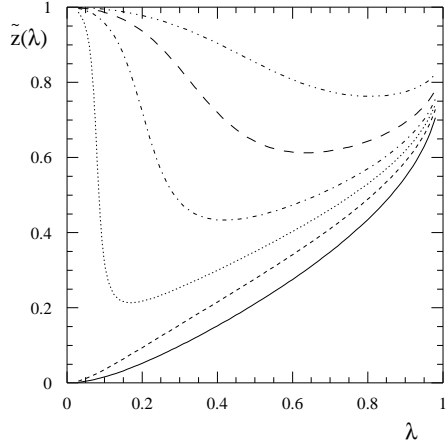,width=8cm,height=8cm}
    \caption{Solution of the eigenvalue equation in the limit 
      ${\cal N}\to\infty$ in the Euler-McLaurin approximation
      for different values of $\alpha$.
      Solid line: $\alpha=0$, dashed line: $\alpha=0.2$, dotted 
      line: $\alpha=0.35$, dash-dotted line: $\alpha=0.5$, long-dashed
      line: $\alpha=0.8$, dot-dot-dashed line: $\alpha=1.5$.
      The exact results are not displayed as they
      almost coincide with the ones in the figure.}
    \label{fig:5}
  \end{center}
\end{figure}

The numerical solutions of Eq.~(\ref{eq:4.030}) 
are displayed in Fig.~\ref{fig:5} for different values of $\alpha$.
It is interesting to follow the behavior with $\alpha$ of
the curves, from the $\alpha\to 0$ case (strong coupling), 
which still carries the fingerprints of the harmonic oscillator, to the 
$\alpha\to\infty$ one (weak coupling), which corresponds to the straight 
line $\tilde z=1$. 

The behavior of the curves 
is ruled both by the function  $\xi_\infty(\lambda)$,
which goes to $-\infty$ when $\lambda\to 1$, whereas for $\lambda\to 0$ 
\begin{equation}
  \label{eq:9.303}
  \xi_\infty(\lambda)\stackrel{\lambda\to0}\longrightarrow
\left\{  \begin{array}{cc}
    +\infty & {\rm for ~}\alpha<\alpha_{\rm cr}\\
    -\infty & {\rm for ~}\alpha>\alpha_{\rm cr}
    \end{array}
\right.
\ \ \ \ \ \ \mbox{with}\ \ \ \ 
\alpha_{\rm cr}=\frac{3}{8}\sqrt{\frac{3}{5}}\simeq 0.29\ ,
\end{equation}
and by the monotonic decrease of $\phi(\tilde z)$ 
in the interval $(0,1)$, at whose endpoints it assumes the values
\begin{equation}
  \label{eq:9.305}
  \lim_{\tilde z\to 0}\phi(\tilde z)=
   +\infty\qquad \mbox{and}\qquad\lim_{\tilde z\to 1}\phi(\tilde z)=-\infty~.
\end{equation}
Thus, when $\lambda\to1$,
\begin{equation}
\tilde z(\lambda)\to 1+\frac{1}{4\log(1-\lambda)}
\end{equation}
no longer depends upon $\alpha$ and  all 
the curves in Fig.~\ref{fig:5} coalesce to 1.
Indeed since the pairing interaction is of finite range, a pair trapped in
highly excited harmonic oscillator states has the
two partners sufficiently de-localized to be little affected by
the interaction. 
In this connection it is worth reminding that the classical limit is achieved
by letting the degeneracy of the single particle levels become very large
\cite{Roman:2002dh}. Thus the coincidence of all the eigenvalues, for any $G$, in
$\lambda$=1 also reflects the evolution from quantum to classical
mechanics of our system.

For $\lambda\to 0$ two cases occur: if $\alpha<\alpha_{\rm cr}$ 
then $\tilde z(\lambda)\to 0$, whereas if $\alpha>\alpha_{\rm cr}$ then
$\tilde z(\lambda)\to 1$. Thus a transition occurs at
$ \alpha_{\rm cr}$, as illustrated in Fig.~\ref{fig:5}:
the almost parabolic behavior of $\tilde z(\lambda)$ for small $\alpha$
(which was first observed in Ref.~\cite{Barbaro:2001ir})
is strongly distorted for $\alpha \simeq \alpha_{\rm cr}$ for small $\lambda$.
For larger $\alpha$ a smoother behavior is recovered. In particular in 
Fig. \ref{fig:5} a marked minimum is seen to develop for $\alpha$
above, but close to, the critical value.

The convergence of all the curves of Fig.~\ref{fig:5} to
$\tilde z(0)=0$ for $\alpha<\alpha_{\rm cr}$ reflects the pressure 
exercised by the infinite number
of the high-lying, large degeneracy, levels on the low-lying, low-degeneracy, 
ones.
This occurrence might be understood on the basis of a sum rule the trapped 
solutions should fulfill and of the nature of the pairing force. 
Actually the sum of the eigenvalues $\tilde z(\lambda)$
can be exactly computed in the two limiting
cases $\alpha=0$ and $\alpha=\infty$: from the Vi\`ete equations one indeed
has
\begin{equation}
  \label{eq:xxx}
  \Sigma(\alpha=0) 
\equiv \frac{1}{{\cal N}-1} \sum_{k=1}^{{\cal N}-1} \tilde z(k) 
= \frac{1}{4}
\end{equation}
and 
\begin{equation}
\Sigma(\alpha=\infty)=1~,
\end{equation}
respectively. These values set the limits 
for the area under the curves of Fig.~\ref{fig:5}.
The existence of a critical point reflects the fact that,
since the action of the pairing force is gauged by the product $G\Omega_k$, 
above some critical value of $\alpha$ the system prefers to obey the sum 
rule by lifting the lowest eigenvalues (corresponding to the lowest 
degeneracies) to the unperturbed values. Indeed, as we shall illustrate,
if the degeneracy of the single particle levels is not growing fast
enough with $k$, then no transition occurs.

The case of finite $\cal N$ will not be discussed here.
We just mention that, as proved in Ref.~\cite{Barbaro:2003bb}, 
the eigenvalues obtained in the $\cal N\to\infty$ are very robust with
respect to variations of $\cal N$, when $G$ is large: indeed
they keep their validity even for values of $\cal N$ as small as five.

It is now natural to ask whether the transition previously discussed 
is peculiar of the harmonic oscillator model or is more general.
To answer this question, 
we take ${\cal N}$ large and consider two classes of models with equally
spaced single particle energies, as in the harmonic oscillator case,
but degeneracies growing (model $a$) or decreasing (model $b$) with
$k$, namely
\begin{eqnarray}
  \label{eq:4.3.2}
  \Omega_k^{(a)} &=& (k+1)^{\gamma}
 \simeq {\cal N}^{\gamma} \lambda^{\gamma}\\
 \Omega_k^{(b)} &=&
\left({\cal N}-k\right)^\gamma
 = {\cal N}^{\gamma} (1-\lambda)^{\gamma}~,
\end{eqnarray}
with $\gamma \ge 0$. The associated expansion parameters are then
\begin{equation}
  \label{eq:4.3.5}
  \alpha^{(a)} = \alpha^{(b)} = \sqrt{\frac{(\gamma+1)^3}{(\gamma+3)}}
\frac{2}{(\gamma+2)\tilde G {\cal N}^{\gamma}}\ .
\end{equation}
Note that for $\gamma$=2 and ${\cal N}$ large the harmonic oscillator 
expression for $\alpha$ is recovered, up to a factor of two.

By performing again the Euler-McLaurin approximation and studying the
behavior of the trapped solution a transition is found to occur
only for the models $a$, with
\begin{equation}
  \label{eq:4.3.8}
  \alpha_{\rm cr}^{(a)}
  =\sqrt{\frac{(\gamma+1)^3}{\gamma+3}}\frac{1}{\gamma(\gamma+2)} ~,
\end{equation}
clearly behaving as $1/\gamma$ when $\gamma\to 0$. 
On the contrary, for the models $b$ no transition can take place.
Thus a transition occurs only if the degeneracy grows with $k$, and
the `strong coupling' domain becomes wider ($\alpha_{\rm cr}$
increases) as $\gamma$ approaches zero: here the transition disappears.
If the degeneracy decreases with $k$ (case b), no transition exists.

\subsection{Two pairs}

We now address the problem of two pairs living in many levels, which is 
already quite instructive on the general case of many pairs and on the 
evolution towards the BCS limit.

In the previous Section we have shown that even when the system consists
of only one pair a transition between two different regimes is found, a
precursor of the quantum phase transition occurring in infinite systems.
When the system is made of several pairs, this phenomenon becomes more 
complicated, since many critical values of the coupling $G$ arise.
These critical values correspond to a particular ``escape'' mechanism of the 
trapped solutions from the grid of the single particle levels, which is 
necessary for the occurrence of a fully collective state, as we shall 
now illustrate.


As we shall see, as for the one pair case, a transiton between two different 
regimes still occurs, but now the number of critical points can be,
under suitable conditions, two, due to the larger number of
possible configurations. If the number of single particle levels 
tends to infinity the two critical values of $G$ merge into one, which, 
notably, coincides with the $G_{\rm cr}$ of a one pair system.
In correspondence of this $G_{\rm cr}$ the system undergoes
a transition from a mean field to a pairing dominated regime.

The Richardson equations (\ref{eq:system}) reduce, for $n=2$, to the following 
system of two equations
\begin{equation}
\left\{
\begin{array}{ccc}
1-G\sum\limits_{\mu=1}^{L} \dfrac{\Omega_{\mu}}{2e_{\mu}-E_1} + 
\dfrac{2G}{E_2-E_1} &=&0\\ 
1-G\sum\limits_{\mu=1}^{L} \dfrac{\Omega_{\mu}}{2e_{\mu}-E_2} + 
\dfrac{2G}{E_1-E_2} 
&=&0~.
\end{array}
\right.
\label{sistema}
\end{equation}
As discussed in Section~\ref{sec:nondeg} the solutions are classified in terms
of like-seniority: we shall ascribe the value $v_l$= 0 to the fully collective state,
$v_l$=2 to a state set up with one trapped pair energy and one
displaying a collective behavior and $v_l$=4 to the state with
two trapped pair energies.

Let us first consider the weak coupling limit, where of course no collective 
mode develops (hence $v_l$ has no significance).
Adopting a perturbative treatment for the pair energies
$E_1$, $E_2$ we write
\begin{equation}
E_i= 2 e_{\mu_i} + G x_i~,
\end{equation}
$G x_i$ being a perturbation. Then
\begin{equation}
\sum_{\mu}\dfrac{\Omega_\mu}{2 e_\mu -E_i}= 
-\dfrac{\Omega_{\mu_i}}{G x_i} + \sum_{\mu \ne \mu_i} 
\dfrac{\Omega_\mu}{2(e_{\mu}- e_{\mu_i})- G x_i}
\end{equation}
and, expanding in $G$, the system \eqref{sistema} becomes 
\begin{equation}
\left\{
\begin{array}{ccc}
1+ \dfrac{\Omega_{\mu_1}}{x_1}-G\sum\limits_{\mu\ne \mu_1} 
\dfrac{\Omega_{\mu}}{2(e_{\mu}-
e_{\mu_1})} + 
\dfrac{2G}{2(e_{\mu_2}-e_{\mu_1})} + O(G^2)&=&0 \\
1+ \dfrac{\Omega_{\mu_2}}{x_2}-G\sum\limits_{\mu\ne \mu_2} 
\dfrac{\Omega_{\mu}}{2(e_{\mu}-
e_{\mu_2})} + 
\dfrac{2G}{2(e_{\mu_1}-e_{\mu_2})} + O(G^2)
&=&0\,,
\end{array}
\right.
\label{sistema2bis}
\end{equation}
where the indices $\mu_1$ and $\mu_2$ select one 
configuration out of the unperturbed ones.
At the lowest order in $G$ (weak coupling regime), 
if $\mu_1\ne \mu_2$ (namely if the two pairs sit on
different single particle levels when $G=0$), 
one has $x_i=-\Omega_{\mu_i}$ and the pair energies
$E_i= 2e_{\mu_i}-G\Omega_{\mu_i}$ are real. Thus the total energy 
$E=E_1+E_2$ becomes
\begin{equation}
E= 2(e_{\mu_1} + e_{\mu_2}) - G (\Omega_{\mu_1}+ \Omega_{\mu_2})~.
\label{weakdiverso}
\end{equation}
If $\mu_1=\mu_2$, which is possible only if $\Omega_{\mu_1}>1$,
the Richardson system is
\begin{equation}
\left\{
\begin{array}{ccc}
1+ \dfrac{\Omega_{\mu_1}}{x_1}-G\sum_{\mu\ne \mu_1} 
\dfrac{\Omega_{\mu}}{2(e_{\mu}-
e_{\mu_1})} + \dfrac{2}{x_2-x_1} + O(G^2)&=0 \\
1+ \dfrac{\Omega_{\mu_2}}{x_2}-G\sum_{\mu\ne \mu_2} 
\dfrac{\Omega_{\mu}}{2(e_{\mu}-
e_{\mu_2})} + \dfrac{2}{x_1-x_2} + O(G^2)&=0
\label{sistema3}
\end{array}
\right.
\end{equation}
and from its solution 
\begin{equation}
x_{1,2}  = -(\Omega_{\mu_1}-1)\pm i \sqrt{\Omega_{\mu_1}-1} 
\label{sistema3bis}
\end{equation}
the complex conjugate pair energies $E_1=E_2^*$ are obtained. 
The total energy reads then
\begin{equation}
E= 4 e_{\mu_1}  - 2 G (\Omega_{\mu_1}-1)
\label{weakidentico}
\end{equation}
and is, of course, real.

In comparing \eqref{weakdiverso} and \eqref{weakidentico}
with the energy of one pair system in the weak coupling regime,
namely
\begin{equation}
E=2 e_{\mu}-G\Omega_\mu~,
\label{weakuno}
\end{equation} 
one sees that, while \eqref{weakdiverso} corresponds to the sum of two 
contributions like \eqref{weakuno}, hence the two pairs ignore each other, 
in \eqref{weakidentico} a positive (repulsive) energy  $2G$ associated to 
the Pauli blocking appears.

Let us now consider the strong coupling domain.
We shall deal only with the states  $v_l=0$ 
and $2$ since the $v_l=4$ states are of minor physical interest.

The $v_l=0$ state arises from an unperturbed configuration with the two
pairs in the lowest single particle level (if $\Omega_1>1$) 
or in the two lowest ones (if $\Omega_1=1$).

Following the same procedure as in the one pair case,
we introduce the variables 
\begin{equation}
x_i=\frac{E_i-2\bar e}{G\Omega}~
\end{equation}
and solve the system \eqref{sistema} in terms of the expansion parameter $\alpha$ defined
in (\ref{eq:n6.7}) through a recursive linearization, getting 
\begin{eqnarray}
x_{1,2}&=&-\frac{(\Omega -1)}{\Omega}-
\alpha^2\frac{(\Omega-2)}{(\Omega-1)} +\alpha^3 \gamma 
\frac{(\Omega-4)}{(\Omega-1)}\nonumber\\
&\pm& i \frac{\sqrt{\Omega-1}}{\Omega}
\mp i \frac{1}{2} \alpha^2 \frac{\Omega}{(\Omega-1)^{3/2}}
\pm i \alpha^3 \gamma \frac{\Omega}{(\Omega-1)^{3/2}}
+{\cal O}(\alpha^4)\,.
\end{eqnarray}
Note that the two pair energies are always complex conjugate and
the system's total energy reads
\begin{equation}
E=4 \bar e-2 G (\Omega-1)\left[1+\alpha^2 
\frac{\Omega(\Omega -2)}{(\Omega-1)^2}-
\alpha^3 \gamma 
\frac{\Omega(\Omega-4)}{(\Omega-1)^2} +{\cal O}(\alpha^4)\right]~.
\label{energyn2}
\end{equation}
Hence the leading order contribution coincides with the 
degenerate case collective eigenvalue (see Eq.~\ref{eq:E}), namely
\begin{equation}
E_{\rm deg}= 4 \bar e -2 G (\Omega-1)\,,
\end{equation} 
which therefore represents a good estimate when
the spreading of the mean field levels is small with respect to $G \Omega $, 
as it is natural to expect. 

We note that, when $\Omega\gg 1$, \eqref{energyn2} 
becomes just twice the value \eqref{eq:n6.18} of the collective energy 
of one pair of 
nucleons living in $L$ levels in the strong coupling regime.
Moreover, the imaginary
part of $x_1$ and $x_2$ goes to zero as $1/\sqrt{\Omega}$.
Thus, in this limit, the Pauli interaction between the two pairs
vanishes, as expected: the two pairs behave as two 
free quasi-bosons condensed in a level 
whose energy is given by (\ref{eq:n6.18}).

Let us now come to the like-seniority $v_l=2$ states. 

If the coupling term (expressing the Pauli principle)
in the Richardson equations \eqref{sistema} were absent, then the eigenvalues 
of the $v_l=2$ states could be simply obtained by adding the collective energy $E_1$
carried by one pair and the trapped energy $E_2$ carried by the other pair.

This situation is recovered in the very strong coupling limit,
where all the single particle energies become essentially equal to 
$\bar e$ and both $\bar e$ and $E_2$ are very small
with respect to $E_1$. 
The first equation of the system \eqref{sistema} then becomes
\begin{equation}
\frac{1}{G} + \frac{\Omega}{E_1} -\frac{2}{E_1}=0
\end{equation}
yielding 
\begin{equation}
E_1= -G(\Omega-2)~,
\end{equation}
namely the energy of the state with two pairs and $v=2$ 
in the one level problem. 
This result sets a correspondence between states with $v_l=2$ and $v=2$,
thus connecting {\em seniority} and {\em ``like-seniority''} (or the physics of a
`broken' and a `trapped' pair). 

In the strong coupling domain however the Pauli term in \eqref{sistema} cannot 
be neglected, but is well approximated by $2/\Omega$: hence the system
\eqref{sistema} decouples and can be recast as 
\begin{equation}
  \label{eq:49}
\left\{
\begin{array}{ccc}
  \dfrac{1}{G_{\rm eff}^{(1)}}-\sum_\mu\dfrac{\Omega_\mu}
  {2e_\mu-E_1}&=&0\\
  \dfrac{1}{G_{\rm eff}^{(2)}}-\sum_\mu\dfrac{\Omega_\mu}
  {2e_\mu-E_2}&=&0\,,
\end{array}
\right.
\end{equation}
where
\begin{eqnarray}
  \label{eq:50a}
  \frac{1}{G_{\rm eff}^{(1)}}
\equiv\frac{1}{G}+\frac{2}{E^{(\nu)}_2-E^{(1)}_1}
\simeq\frac{1+\dfrac{2}{\Omega}}{G}
\\
\label{eq:50b}
  \frac{1}{G_{\rm eff}^{(2)}}
\equiv\frac{1}{G}-\frac{2}{E^{(\nu)}_2-E^{(1)}_1}
\simeq\frac{1-\dfrac{2}{\Omega}}{G}
\,.
\end{eqnarray}
Therefore the Pauli principle in the large $G$ regime 
just re-scales the coupling constant in such a way that the
pairing interaction is quenched 
for the collective state and enhanced for the trapped ones.

The results obtained in the strong coupling domain are in agreement with the findings of
Ref.~\cite{Yuz03}, where the problem of the pairing Hamiltonian
for small superconducting grains is studied. The main difference with
the present approach is that there the single particle levels
are non-degenerate: hence the results of Ref.~\cite{Yuz03} for two pairs
can be recovered from ours by setting $\Omega_\nu=1$.

Let us now discuss the existence of one or more 
critical values for the coupling constant $G$.

In the {\it weak coupling regime}, when a state evolves from an unperturbed 
one with the two pairs living in the same level, then the pair
energies $E_1$ and $E_2$ are always {\em complex conjugate}.
On the other hand when the state evolves from an unperturbed one 
having the two pairs living in two different levels, then 
$E_1$ and $E_2$ are {\em real}. 

By contrast, in the {\it strong coupling regime}
the pair energies $E_1$ and $E_2$ 
of the $v_l=0$ state are always complex conjugate.
It is then clear that, if the degeneracy $\Omega_1$ of the lowest single particle level
is greater than one, then the pair energies
$E_1$ and $E_2$ are complex conjugate in both the
weak and strong coupling regime and their behavior with $G$
is smooth.

On the other hand if $\Omega_1=1$, since in the weak coupling limit 
the two pairs must live on different levels, $E_1$ and $E_2$ are necessarily 
real in a neighbourhood of the origin, but become complex in the strong 
coupling regime.
Thus a singularity in their behavior as a function of the
$G$ is bound to occur. 

In this second case it appears natural to surmise that the
singularity takes place when $E_1$ and $E_2$ coincide.
In fact in this case the Pauli term of the Richardson equations diverges
(reflecting the attempt of four particles to sit in one level
that can host only two of them),
and it must be compensated by a divergence in the sum entering into the 
system (\ref{sistema}): this can only happen if $E_1$ or $E_2$
coincides with an unperturbed eigenvalue.

\begin{figure}[ht]
  \begin{center}
    \epsfig{file=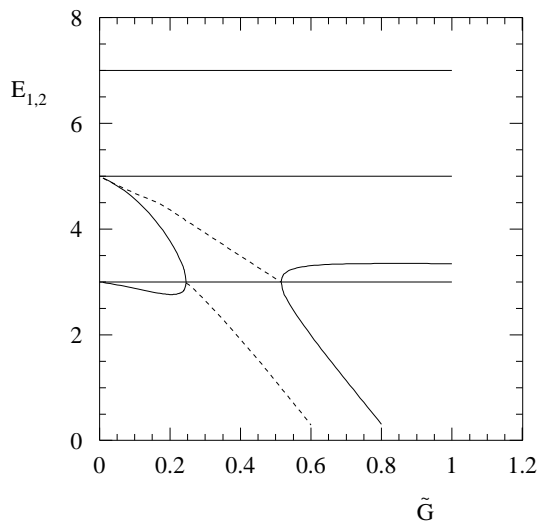,height=9cm,width=9cm}
    \caption{
Behavior with $\tilde G$ of the pair energies of the ground ($v_l=0$)
and first excited ($v_l=2$) states obtained as solutions of the Richardson
equations for a harmonic oscillator well for $L=3$:
dashed lines denote
the common real part of the two pair energies when they
are complex conjugate and solid lines the real part of the two pair energies
when they are different.
All energies are in units of $\hbar\omega_0$. As explained in the text
the two critical points refer to the $v_l=0$ ($\tilde G\simeq 0.21$) and  
$v_l=2$ ($\tilde G\simeq 0.53$) states.
}
    \label{fig:2}
  \end{center}
\end{figure}

The situation is portrayed in
fig.~\ref{fig:2}, where the behavior of the pair energies with 
$\tilde G$ is displayed for a harmonic oscillator well assuming $L=3$.

Clearly in this case two values of $\tilde G_{\rm cr}$ occur.
The pair energies $E_1$ and $E_2$, real in the weak coupling limit, 
coincide at the critical point $\tilde G_{\rm cr}^{(1)+}$ 
(their common value being $2 e_1$)
and then become complex conjugate; the energy of the
associated state evolves in the $v_l=0$ collective mode. 
By contrast, for the $v_l=2$ state, arising from the configuration with two 
pairs living on the second level (which is allowed for 
the harmonic oscillator well), the two pair energies
$E_1$ and $E_2$ are complex conjugate in the weak coupling limit, coalesce
into the energy $2 e_1$ at the critical point $\tilde G_{\rm cr}^{(1)-}$ and then 
become real. 
One of the two solutions remains trapped above $2 e_1$, while the other 
evolves into a collective state: the sum of the two yields the energy
of the $v_l=2$ state.  
   
Analytic expression of the critical values of $G$ can be obtained starting
from the Richardson equations and assuming $\Omega_\nu=1$.
Indeed in Ref.~\cite{Barbaro:2003bb} the expression  
\begin{equation}
  \label{eq:29n}
  G_{\rm cr}^{(\nu) \pm}=\frac{1}{{\cal P}_{(1)\nu}\pm
    {\cal P}_{(2)\nu}}=\left[\sum_{\mu=1(\mu\neq\nu)}^L 
    \frac{\Omega_\mu}{2 e_\mu-2 e_\nu}\pm 
    \sqrt{\sum_{\mu=1(\mu\neq\nu)}^L
    \frac{\Omega_\mu}{(2 e_\mu-2 e_\nu)^2}}\,\right]^{-1}~,
\end{equation}
where $G^+$ is the critical value of the coupling related to the ground
state and $G^-$ the one related to the excited states, is derived.
In the above the quantities
\begin{equation}
  \label{eq:13n}
  {\cal P}_{(k)\nu}=\left\{\sum_{\mu=1(\mu\neq\nu)}^L
\frac{\Omega_\mu}{(2 e_\mu -2 e_\nu)^k}  \right\}^{\frac{1}{k}}
\end{equation}
are the inverse moments of the level distribution.
We thus recover, although through a somewhat different route,
the results found long ago by Richardson~\cite{Rich}.

It is important to stress that 
for the occurrence of a critical value of $G$ 
a single particle level  with $\Omega=1$ {\em must exist}.
Such level is the lowest one in a harmonic oscillator: for this 
potential well both the ground and the first excited state of a  
$n=2$ system carry one critical value of the coupling constant, namely
$G_{\rm cr}^{(1)+}$ and $G_{\rm cr}^{(1)-}$, respectively.
In correspondence of these values of $G$ the pair energies take on the
value $E_1=E_2=2e_1$. Thus for a $n=2$
system two (at most) critical points exist on a $\Omega=1$ level.

\begin{figure}[ht]
  \begin{center}
    \epsfig{file=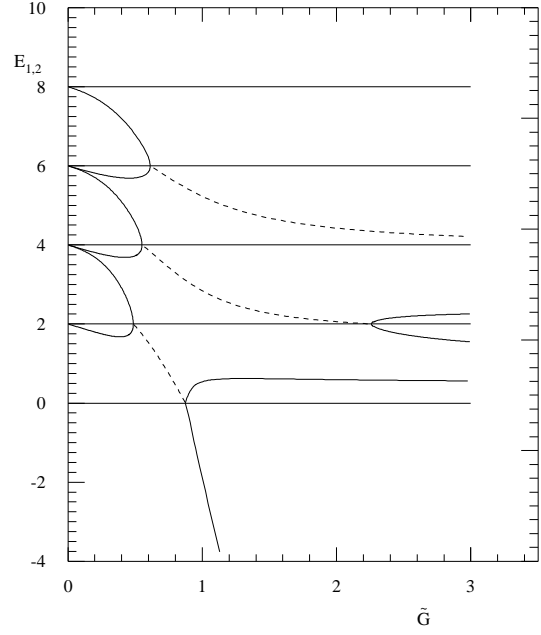,height=10cm,width=10cm}
    \caption{
Behavior of the pair energies of three excited states
in the case of $\Omega_\mu=1$. The number of levels considered is 
$L=9$. The energies $E_{1,2}$ are expressed in units of the level spacing.}
    \label{fig:4}
  \end{center}
\end{figure}

Finally, owing to the relevance of the $\Omega=1$
degeneracy, we consider the model of $L$, for simplicity equally spaced, 
single particle levels all having 
$\Omega=1$,  a situation occurring in metals and in deformed nuclei.
In this instance two positive $G_{\rm cr}$ always exist in the lowest level
when $L\geq 3$ (in fact $G_{\rm cr}^{(1)-}\to\infty$ for $L=2$).
Moreover a positive $G_{\rm cr}^{(1)-}$ implies complex $E_1$ and $E_2$ for
$G<G_{\rm cr}^{(1)-}$ and, since for small $G$ both the pair energies are real,
they should evolve from an excited unperturbed configuration having the two
pairs placed in the two lower-lying single particle levels,
as illustrated in fig.\ref{fig:4}.
Numerically, for this model, 
we have found that two $G_{\rm cr}$ appear on the second level when 
$L\ge 9$ and on the third level when $L\ge 16$.
Thus in the $\Omega=1$ model two pairs can form, so to speak, a 
{\em quartet} only if they live on adjacent single particle levels
in the unperturbed configuration. Furthermore the more excited the unperturbed
configuration is
the more not only $G$, but also $L$, should be larger for the merging to 
occur, a fact clearly reflecting the competition between the mean field
and the pairing force.
Finally we observe that here, at variance
with the finding of Ref.~\cite{Roman,Roman:2002dh,Hase},
$G_{\rm cr}^{(\nu +1)+}> G_{\rm cr}^{(\nu)+}$:
this is simply because we measure the single particle energies from the bottom of the well
rather than from the Fermi surface.

It is  interesting to establish the relationship of the above discussed
critical values of $G$ with the critical value \eqref{eq:9.303} found for
one pair, which corresponds to
\begin{equation}
\label{gcrho}
G_{\rm cr}\simeq \frac{8}{L^2}
\end{equation} 
when $L$ is large. 
For the $n=2$ system we first notice that, at large $L$, in 
$G_{\rm cr}^{(1)\pm}$
the moment ${\cal P}_1$ dominates over ${\cal P}_2$: 
hence, in this condition, 
\begin{equation}
G_{\rm cr}^{(1)+}\simeq G_{\rm cr}^{(1)-}~.
\end{equation}
Moreover, again for $L$ large enough, 
\begin{equation}
{\cal P}_1\simeq \frac{L^2}{8}~,
\end{equation}
entailing the equality
\begin{equation}
G_{\rm cr}(n=1)=G_{\rm cr}(n=2)
\end{equation}
in the asymptotic $L$ limit. 
From this outcome one sees that also when $n=2$ the relevant dynamical 
element for $G<G_{\rm cr}^{(1)+}$ is the mean field, whereas for 
$G>G_{\rm cr}^{(1)-}$ it is the pairing force, as far as the system's ground 
state is concerned. 

\subsection{Transition amplitudes}
\label{sec:trans}

Further insight into the critical behavior of the 2-pair system
can be gained by studying the pair transfer matrix elements from a one
pair to a two pairs state as a function of $G$.

This can be achieved by constructing the wave functions according to
Eqs.~\eqref{eq:1}, \eqref{psin}, \eqref{eq:beta} with $n=1$ and 
$n=2$ and, with these, 
calculate the transition amplitudes induced by the operator $\hat A^\dagger$ 
(see Eq.~\eqref{eq:A}), namely the matrix elements
\begin{equation}
<n=1|\hat A^\dagger|n=0>\ \ \ \mbox{and}\ \ \ 
<n=2|\hat A^\dagger|n=1>~.
\label{eq:tme}
\end{equation}

In the fully degenerate case the general expression for these transition
amplitudes between states with any number $n$ of pairs and seniority $v$,
can be obtained using the expression (\ref{eq:6}) for the excitation spectrum.
Indeed, inserting into the pairing spectrum a complete 
set of states $|n',v'>$, one has
\begin{equation}
\sum_{n',v'}  <n+1,v|\hat A^\dagger|n',v'> 
<n',v'|\hat A|n+1,v>
=\left(n+1-\frac{v}{2}\right)
    \left(\Omega-n-1-\frac{v}{2}\right)\,,
  \label{eq:degsp1}
\end{equation}
thus getting for the transition amplitude
\begin{equation}
  \label{eq:ax3}
  <n+1,v|\hat A^\dagger|n,v>=\sqrt{\left(n+1-\frac{v}{2}\right)
    \left(\Omega-n-\frac{v}{2}\right)}\,.
\end{equation}
Note the $G$-independence of \eqref{eq:ax3}, which, moreover, is diagonal
in the seniority quantum number.
From the previously illustrated analogy between seniority and like-seniority,
we expect that
for large $G$ the matrix elements \eqref{eq:tme} display
an asymptotic behavior coinciding with \eqref{eq:ax3}, namely
\begin{eqnarray}
  \label{eq:ax11}
  <1,v_l=0|\hat A^\dagger|0>~&\xrightarrow[G\to\infty]{}&~
  \sqrt{\Omega}\\
  <2,v_l|\hat A^\dagger|1,v_l>~&\xrightarrow[G\to\infty]{}&~
  \begin{cases}
    \sqrt{2(\Omega-1)}&\text{for~}v_l=0\\
    \sqrt{\Omega-2}&\text{for~}v_l=2~.
  \end{cases}
\end{eqnarray}
Actually this happens only if $v_l$ is conserved. 
However, the number $v_l$ is not
conserved at finite $G$: only in the strong coupling regime the 
vanishing of the matrix element between states of different Gaudin numbers
occurs.
Indeed for finite $G$ we get
\begin{multline}
  \label{eq:ax5}
  <(m_1,m_2)|\hat A^\dagger|(m)>=\\
  \frac{1}{\cal N}\sum_{\mu\nu}
  [\tilde \beta^{(1)}_\mu(m_1,m_2)\tilde \beta^{(2)}_\nu(m_1,m_2)]^*
  \left[\tilde \beta_\mu(m)\sqrt{\Omega_\nu}
    +\tilde \beta_\nu(m)\sqrt{\Omega_\mu}
  -\dfrac{2}{\sqrt{\Omega_\mu}}\beta_\mu(m)\delta_{\mu\nu}\right]~,
\end{multline}
${\cal N}$ being a normalization constant.

Using the Richardson equations to get rid of the sums, we recast 
\eqref{eq:ax5} as follows
\begin{equation}
  \label{eq:ax51}
   <(m_1,m_2)|\hat A^\dagger|(m)>=
   \frac{1}{{\cal N}}
   \frac{2C_1(m_1,m_2)C_2(m_1,m_2)C_1(m)}{G[(E_1(m_1,m_2)-E(m)]
[(E_2(m_1,m_2)-E(m)]}~,
\end{equation}
which is real since $E_1$ and $E_2$ are either real or complex 
conjugate.

\begin{figure}[htp]
  \begin{center}
        \mbox{
      \begin{tabular}{cc}
        \epsfig{file=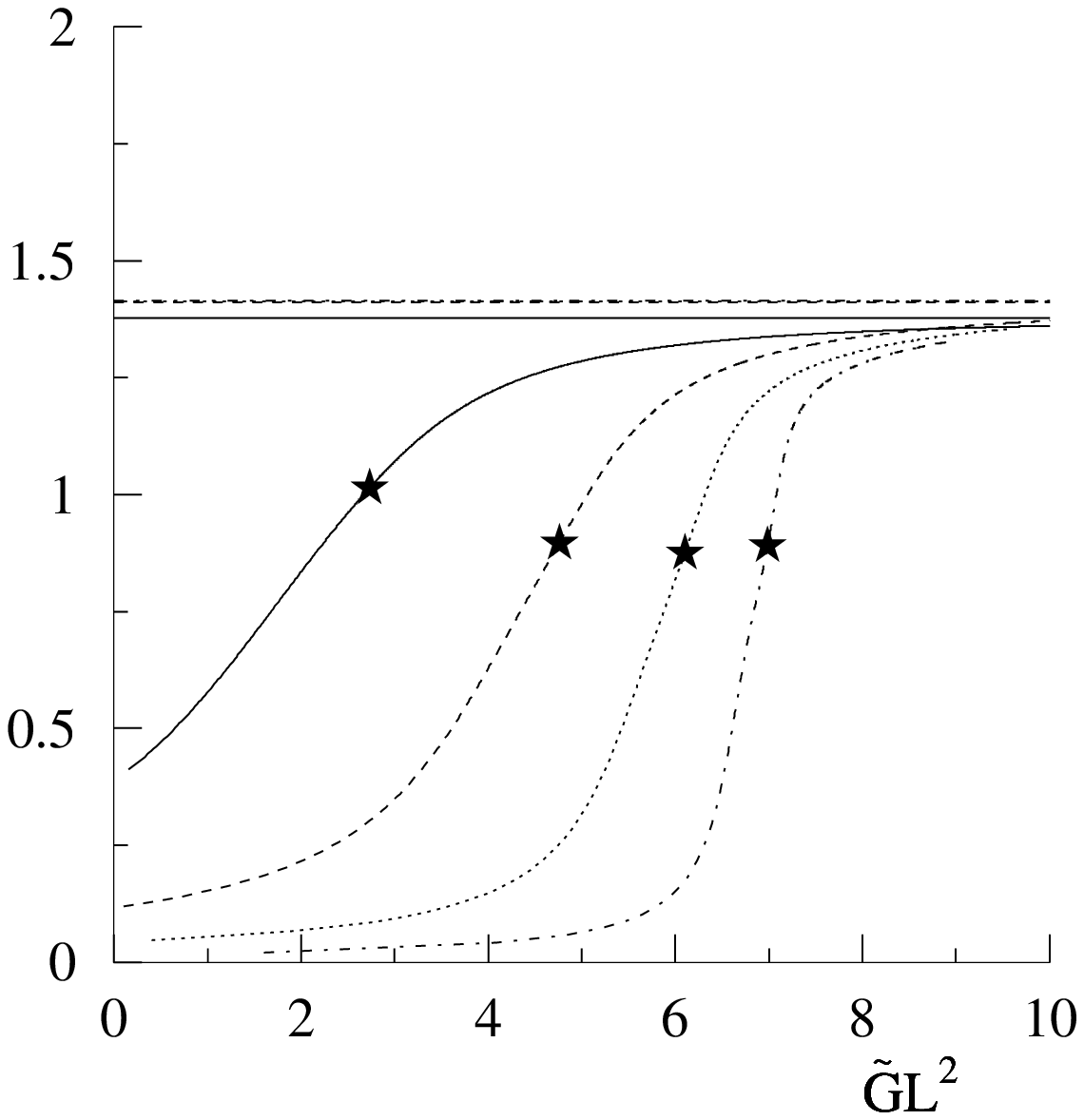,width=6cm,height=7cm}&
        \epsfig{file=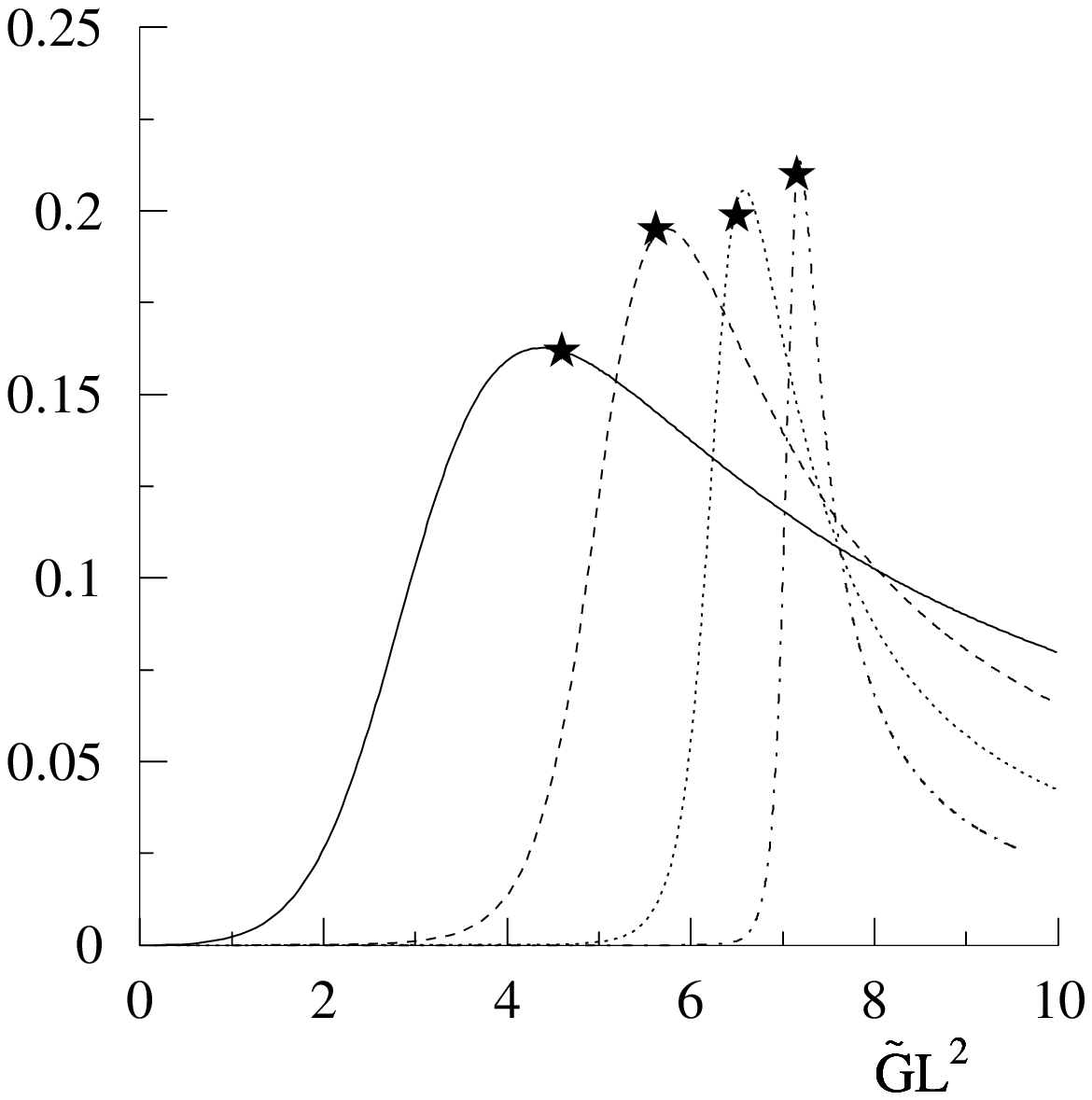,width=6cm,height=7cm}
        \\
        \epsfig{file=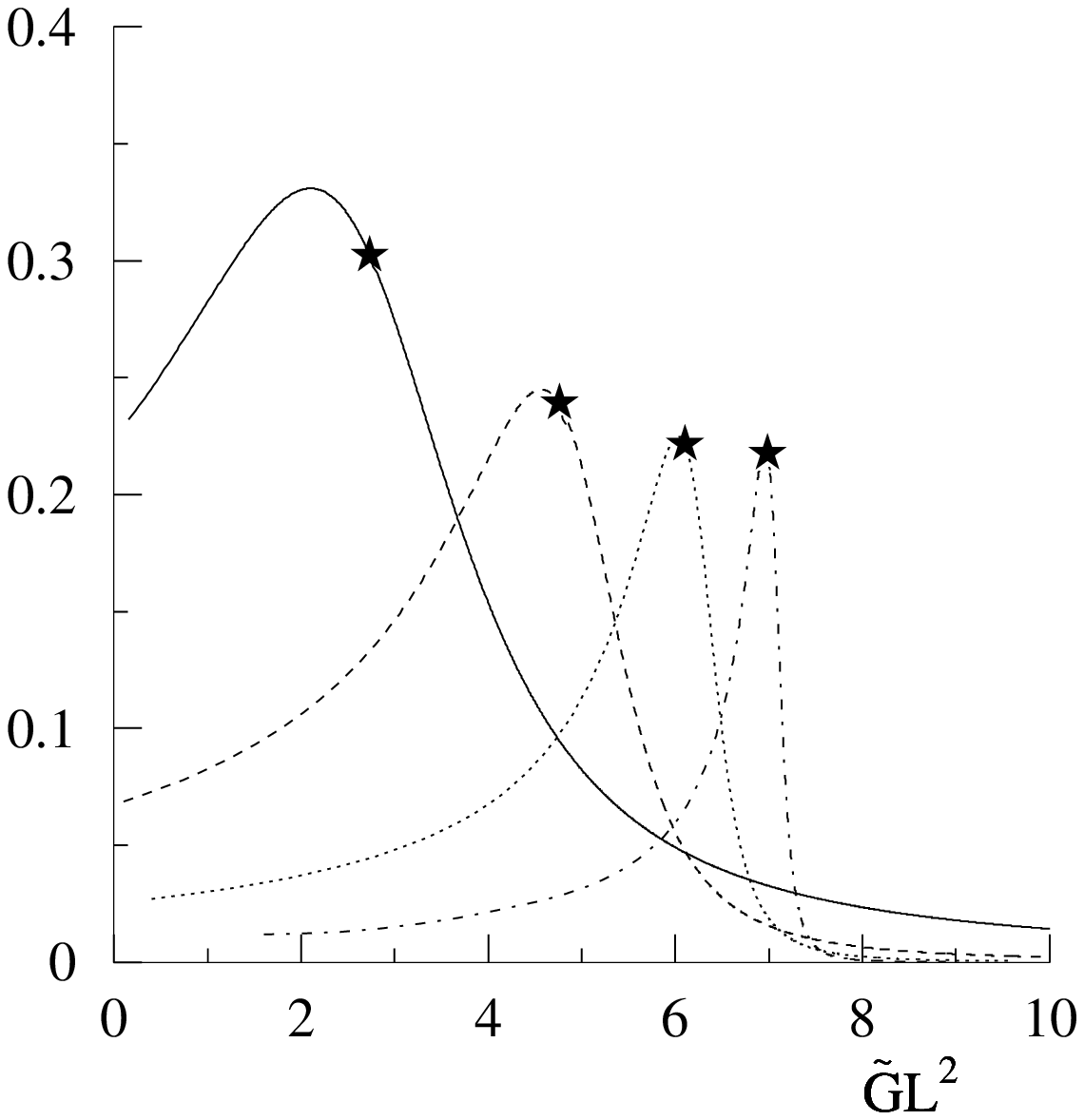,width=6cm,height=7cm}&
        \epsfig{file=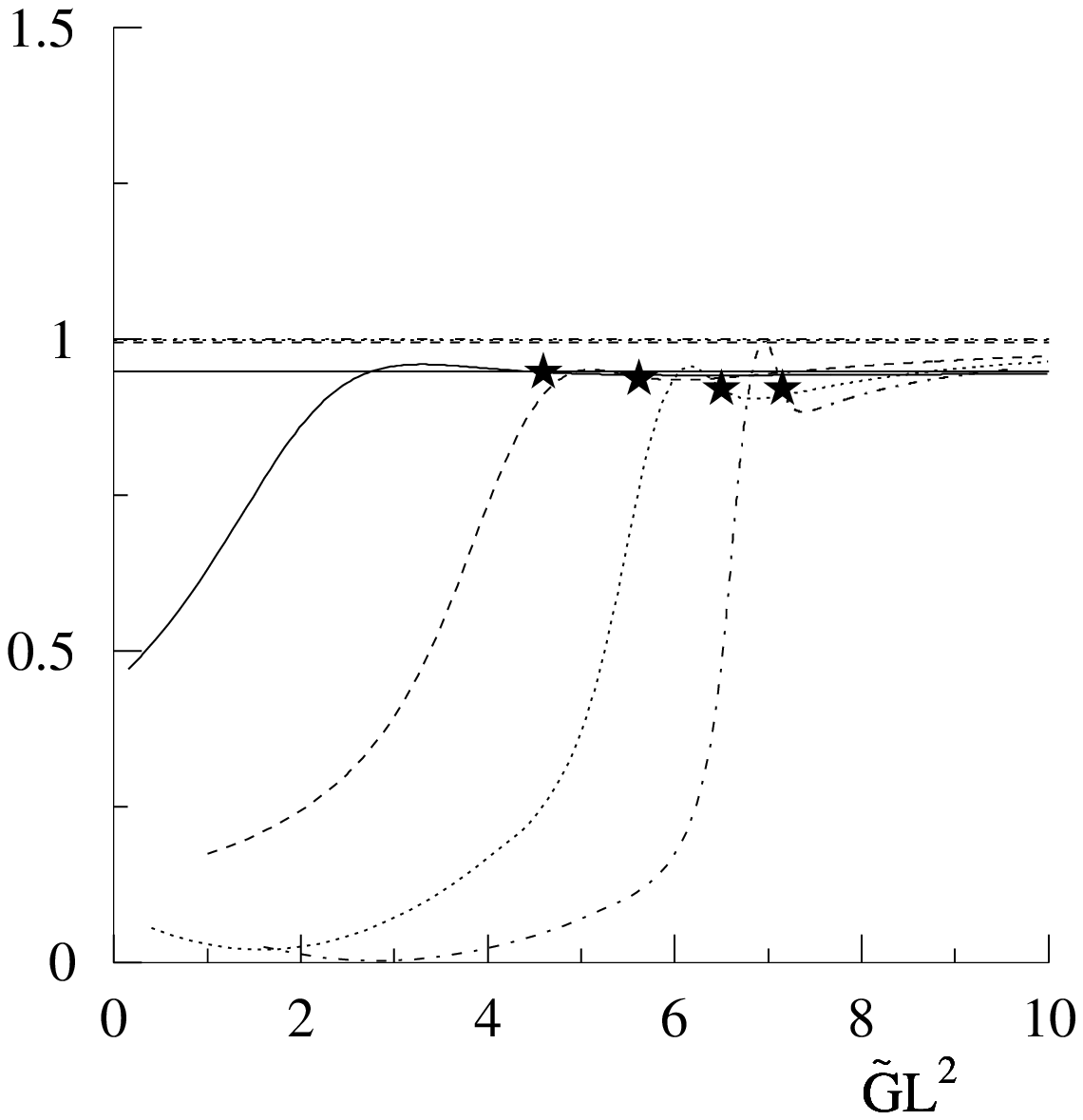,width=6cm,height=7cm}
        \end{tabular}
      }
    \caption{The transition matrix elements \eqref{eq:ax51} divided by 
     $\sqrt{\Omega}$ for $L=$4 (solid), 10 (dashed), 20 (dotted) and 40
      (dot-dashed) levels of a h.o. well.  
      Upper left panel: transition from the $m=1$ (collective)
      state to the $(m_1,m_2)=(1,2)$ ($v_l=0$) state;
      upper right panel: transition from the $m=1$ (collective)
      state to the $(m_1,m_2)=(2,2)$ ($v_l=2$) state;
      lower left panel: transition from the $m=2$ (trapped)
      state to the $(m_1,m_2)=(1,2)$ ($v_l=0$) state;
      lower right panel: transition from the $m=2$ (trapped)
      state to the $(m_1,m_2)=(2,2)$ ($v_l=2$) state.
      An asterisk denotes the position of the critical points.
      The straight lines represent the asymptotic values for $G\to\infty$.
}
    \label{fig:7}
  \end{center}
\end{figure}
We display in 
fig.~\ref{fig:7} the amplitudes \eqref{eq:ax51} 
for the transfer from a $n=2$ to a $n=1$ system
(divided, for obvious convenience, by $\sqrt{\Omega}$).
We consider  
the $n=1$ system either in the ground collective state 
($m=1$) or in the first excited trapped state ($m=2$).
On the other hand the $n=2$ system is either in the $v_l=0$
(namely $(m_1,m_2)=(1,2)$) or in the $v_l=2$ ($(m_1,m_2)=(2,2)$) state.
The calculation is performed for a harmonic oscillator well with $L=4,10,20$ and $40$.

The matrix elements clearly exhibit a conspicuous
enhancement in the proximity of $G_{\rm cr}$ 
(marked by an asterisk in the figure), which most clearly reflects 
the onset of an ODLRO (off-diagonal-long-range-order) in the system.

The behavior of the transition matrix element around the critical points 
is quite delicate since both the numerator and the denominator in 
\eqref{eq:ax5} tend to vanish when $E_1$ and $E_2$ tend to $2 e_1$.
In this limit the transition amplituse \eqref{eq:ax5} reads
\begin{equation}
  \label{eq:bx21}
  <(m_1,m_2)|\hat A^\dagger|(m)>~\xrightarrow[G\to G_{\rm cr}^\pm]{}~
  \dfrac{2}{G_{\rm cr}^\pm}
  \frac{\sum_{\mu\not=1}
\dfrac{\sqrt{\Omega_\mu}\tilde 
      \beta_\mu(m)}{2\epsilon_\mu-2\epsilon_1}
\pm\tilde \beta_1(m)
    {\cal P}_{(2)}}
  {\sqrt{6{\cal P}_{(2)}^4\pm 8{\cal P}_{(2)}{\cal P}_{(3)}^3
      +2{\cal P}_{(4)}^4}}~.
\end{equation}
Now, the inverse moments of the level distribution \eqref{eq:13n}
vanish in the $L\to\infty$ limit: hence \eqref{eq:bx21} diverges.
As above mentioned, this occurrence, not appearing in fig.~\ref{fig:7} because 
of the chosen normalization, relates to the ODLRO which sets in into a 
system close to a phase transition.

Specifically the diagonal amplitudes for large $L$ behave according to
\begin{eqnarray}
<(m_1,m_2)|\hat A^\dagger|(m)>
\stackrel{L\to\infty} \simeq
  \begin{cases}
\sqrt{2\Omega}\ \theta\left(\tilde G-\tilde G_{\rm cr}\right) \,\,\, v_l=0\\
\sqrt{\Omega}\ \theta\left(\tilde G-\tilde G_{\rm cr}\right) \,\,\,\,\, v_l=2\,.
  \end{cases}
\end{eqnarray}
The off-diagonal amplitudes behave instead as 
\begin{equation}
<(m_1,m_2)|\hat A^\dagger|(m)>
\stackrel{L\to\infty} \simeq
\delta\left(\tilde G-\tilde G_{\rm cr}\right)~.
\end{equation}

From the figures it is also clear that the critical value of $G$ increases
with $L$ and $\tilde G_{\rm cr} L^2\to 8$ as $L\to\infty$ (see Eq.~\eqref{gcrho}).

\section{BCS and the exact solution}
\label{sec:BCS}
\begin{figure}[ht]
 \begin{center}
\def\epsfsize#1#2{0.9#1}
\epsfbox[170 520 450 820]{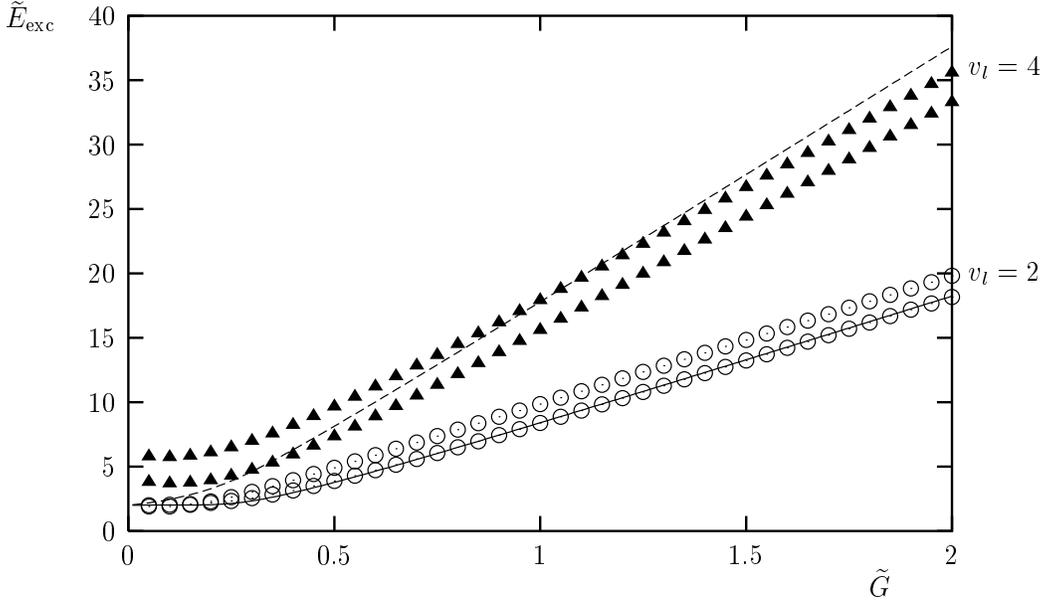}
    \caption{Behavior with $\tilde G$ of the excitation energy
of a two-pairs system in a harmonic oscillator well with $L=3$.
The exact solutions with $v_l=2$ (circles) and $v_l=4$ (triangles) are compared
with the BCS results for $v=2$ (solid line) and $v=4$ (dashed line).
For each like-seniority the two exact solutions develop from different
unperturbed configurations.}
    \label{fig:8}
  \end{center}
\end{figure}
In concluding this work we shortly comment on the relation between 
the exact solution of the 
Richardson equations and the BCS solution in terms of Bogoliubov 
quasi-particles.

To this scope we self-consistently solve the well-known BCS equations for $N$
fermions living in $L$ levels, namely
\begin{eqnarray}
&&v_\nu^2 = \frac{1}{2} \left[1-
\frac{\epsilon_\nu-\mu}{\sqrt{(\epsilon_\nu-\mu)^2+\Delta^2}}\right]
= 1- u_\nu^2\,,
\\
&&\sum_{\nu=1}^L 
\frac{\Omega_\nu}{\sqrt{(\epsilon_\nu-\mu)^2+\Delta^2}}=\frac{2}{G}\,,
\label{lam}
\\
&&\sum_{\nu=1}^L 
\Omega_\nu \left[1-
\frac{\epsilon_\nu-\mu}{\sqrt{(\epsilon_\nu-\mu)^2+\Delta^2}}\right]=N ~,
\end{eqnarray}
where $\Delta= G\sum_\nu\Omega_\nu u_\nu v_\nu$ the gap, 
$\mu$ the chemical potential and 
$\epsilon_\nu= e_\nu-G v_\nu^2$, and we evaluate in this framework
the excitation energy of a system with seniority $v$, 
given by the energy of $v$ quasi-particles, namely
\begin{equation}
E_{BCS}=\sum_{\nu=1}^v \sqrt{(\epsilon_\nu-\mu)^2+\Delta^2}\,.
\label{eq:eQP}
\end{equation}

In fig.~\ref{fig:8} we display and compare the exact excitation energies 
$E_{\rm exc}=E(v_l)-E(g.s.)$ for a $L=3$ harmonic oscillator well and 
the corresponding Bogoliubov's quasi-particles predictions \eqref{eq:eQP} for a
$v=2$ and a $v=4$ state.
It appears that 
for $\tilde G$ larger than the highest critical value ($\tilde G_{\rm cr}^-
\simeq 0.6$) both the BCS and the exact excitation energy become
linear functions of $\tilde G$ and, remarkably, 
are very close to each other, in particular for 
$v_l=2$. 

Note that the excitations in the BCS picture amount to break a pair, 
whereas in the Richardson frame the like-seniority excitations
correspond, microscopically, to two particles-two holes (the $v_l=2$) and to
four particles-four holes (the $v_l=4$) excitations without the breaking
of any pair.
In other words they are associated with the promotion of one or two pairs
to higher lying single particle levels.

The result in Fig.~\ref{fig:8} thus strengthens the correspondence between 
seniority and ``like-seniority''.
Moreover it confirms that the Richardson exact solutions behave as
(\ref{eq:eQP}) in the strong $G$ limit, as was proved by Gaudin
in the large $L$ limit~\cite{Gau95}.
It is remarkable that this appears to be approximately true already for
$L=3$.

\section{Conclusions}

In this report we have presented two approaches to the bosonization of
fermionic systems based on the introduction of composite variables associated
with the relevant degrees of freedom. One relates to the change of variables
in the Berezin integral expressing the generating functional, the other
to the straight solution of the Richardson equations.

Our testing ground has been the simple pairing Hamiltonian, which acts
between fermions coupled to total angular momentum $J=0$. 
Such system exhibits three kinds of excitations, related to the addition of 
removal of pairs, breaking of pairs (the so-called seniority excitations) and
promotion of $J=0$ pairs to higher energy levels, respectively.

The first kind of excitation connects the ground state energies of different
systems, in particular atomic nuclei, 
and is associated with a Goldstone mode in the spectrum.
Although the system we consider is finite, and therefore cannot truly display 
a spontaneous symmetry breaking mechanism (as it happens instead in 
superconductivity), nevertheless the precursor signature of a Goldstone boson
can be clearly identified.
This has been demonstrated, at least in the degenerate case,
by performing a Hubbard-Stratonovitch linearization of
the action and showing that the associated bosonic auxiliary field 
indeed endows the properties of a Goldstone boson. 

The seniority excitations have been studied, again in the degenerate case, 
in the framework of the Hamiltonian
formalism cast in the language of Grassmann variables and a minimal basis 
of composite fields has been set up which
reproduces the full pairing spectrum in the degenerate case together with the
associated wave functions.

When many single particle levels are present, new facets of the pairing
physics disclose themselves. 

First a third type of excitations 
(promotion of pairs to higher levels) can occur. These have been classified
in terms of a new quantum number, the ``like-seniority'', 
closely related to the Gaudin number~\cite{Gau95}, which has been shown
to coincide with the true seniority when the strength of the interaction is
large.
Second, critical values of the coupling constant $G$ appear, signalling the
transition between two different regimes: one dominated by the mean field,
the other by the pairing bosonizing interaction.
Notably the number of the critical values of $G$, which is crucially linked
to the exsistence of a lowest single particle level with pair degeneracy 
$\Omega=1$, increases with the number
of pairs:  however all the $G_{\rm cr}$ collapse into a common value 
when the number of levels tends to infinity.

The next step in the study of the pairing Hamiltonian will be to consider 
the general case of $n$ pairs living in a set of single particle levels and explore 
whether the signatures of a Goldstone boson and of the Higgs excitation 
are still present, so shading further light on the phase transition to 
superconductivity occurring in finite and infinite Fermi systems.
In this connection the escape mechanism of the pairs from the grid of the single particle 
levels should be carefully examined in relation to the pair degeneracy of the latter.

Finally, worth to be explored are new scheme recently proposed for the bosonization of finite
Fermi systems, specifically of atomic nuclei: in particular the one outlined in 
Refs.~\cite{Palumbo:2004yb,Palumbo:2004ab} which, 
always in the framework of the path integral, introduces the concept of
boson dominance in computing the trace expressing the partition function of a nucleus.

\begin{center}
{\bf Acknowledgements}
\end{center}

Most of the work presented here is the fruit of a longstanding collaboration
with R. Cenni, A. Molinari and F. Palumbo, whom we would like to thank for 
many useful discussions and careful reading of the manuscript.

\end{document}